\documentclass[aps,prd,reprint,groupedaddress]{revtex4-2}

\usepackage{amsmath}
\usepackage{amsfonts}
\usepackage{mathrsfs}
\usepackage{amssymb}

\newcommand{\normord}[1]{\;\bm{\vcentcolon}\,\mathrel{#1}\,\bm{\vcentcolon}\;}
\providecommand{\vcentcolon}{\mathrel{\mathop{:}}}

\newcommand{\sgn}{\text{sgn}}

\usepackage{slashed}
\usepackage{simpler-wick}
\usepackage{tensor}
\usepackage{bm}

\begin{document}


\title{Quantum electrodynamics in the null-plane causal perturbation theory II}


\author{O.A. Acevedo}
\email[]{oscar.acevedo@unesp.br}

\author{B.M. Pimentel}
\email[]{bruto.max@unesp.br}
\affiliation{Institute for Theoretical Physics (IFT), S{\~a}o Paulo State University (UNESP), R. Dr. Bento Teobaldo Ferraz 271, S{\~a}o Paulo, SP 01140-070, Brazil}


\date{\today}

\begin{abstract}
We develop a complete formulation of quantum gauge invariance in light-front dynamics for interacting theories with massless vector gauge fields in the framework of null-plane causal perturbation theory. We apply the general results to quantum electrodynamics, showing that the so-called ``gauge terms'' present in the photon commutation distribution when quantized under the null-plane gauge condition have no contribution in the calculation of the physical $S$-operator matrix elements at any order. We use this result to prove the normalizability of the theory, and to calculate the electron's self-energy at second order.
\end{abstract}


\maketitle


\section{Introduction}

Light-front dynamics is one of the three forms of relativistic dynamics discovered by Dirac in 1949 \cite{Dirac}, complemented in 1978 with two additional forms by Leutwyler and Stern \cite{LeutwylerStern}. In this dynamical form, in which the isochronic surfaces are null planes of constant $x^+\propto x^0+x^3$, the number of Poincar{\'e} generators independent of the interaction is maximum \cite{Dirac}, and, in the quantum theory, the vacuum state of the interacting theory is far more simple than that of instant dynamics \cite{Burkardt}. These peculiarities allow the implementation of techniques which are almost impracticable in instant dynamics, as Tamm-Dancoff's truncation \cite{Perry}, which turns light-front field theory into a very useful tool for the study of hadron physics \cite{Igreja}.

In spite of this, the equivalence between the instant and the light-front formulations of quantum field theory is not firmly established yet. This problem for perturbative quantum electrodynamics (QED) was first addressed by Ten Eyck and Rohrlich \cite{TenEyck1,TenEyck2} and by Yan \cite{ChangYan1,ChangYan2}. In their calculations, Feynman's amplitudes at one-loop level exhibited double-pole singularities because of the instantaneous terms that appear in the gauge field propagator when quantized in the null-plane gauge $A^+=0$; this problem was solved by Pimentel and Suzuki \cite{PimentelSuzuki1,PimentelSuzuki2}, who proposed a prescription to treat those poles in a causal way. However, the importance of the instantaneous terms in the fermion and gauge fields propagators is not clear yet; recent reviews on the \textit{status quo} of the gauge field propagator can be found in Refs. \cite{SuzukiSales1,SuzukiSales2}. Also, very recently, the equivalence problem for one-loop radiative corrections was studied in Refs. \cite{BhamreMisra1,BhamreMisra2}, and the fulfilment of Ward-Takhashi's identity at one-loop order in Ref. \cite{SuzukiJi}.

Aiming to shed light to the subtleties of perturbative light-front field theory, the authors developed the framework of null-plane causal perturbation theory (CPT) \cite{APS21,APS22}, an axiomatic approach to the $S$-matrix program initiated by Heisenberg \cite{HeisenbergS} in 1943, and axiomatized in the works by Stückelberg and Rivier \cite{StueckelbergBook, Stueckelberg} and Bogoliubov, Medvedev and Polivanov \cite{BogoMedPoli, Bogo, BogoLogunov}. The detailed perturbative solution to Bogoliubov-Madvedev-Polivanov's axioms in instant dynamics was carried out in 1973 by Epstein and Glaser \cite{EpsteinGlaser}, in a method in which the causality axiom plays an essential role, and its first application to QED was done by Scharf in 1989 \cite{ScharfFQED}. This approach has the advantage of needing no regularization as the distributional character of the quantized fields is considered. Additionally, no Feynman's propagators appear in loop distributions, which in light-front dynamics means that the problems of the double poles previously referred are avoided. This program was successfully applied to obtain the radiative corrections for Yukawa's model \cite{AP2}, directly showing the equivalence with the instant dynamics formulation \cite{ABPS}.

In a previous paper \cite{APQED1}, that started the series of which the present one is the second part, the authors started the study of QED in light-front dynamics in the framework of null-plane CPT, in which the equivalence with instant dynamics was accomplished for the scattering processes and vacuum polarization, \textit{when gauge invariance is taken into account}. More precisely, we have seen that the photon quantized field operator in the null-plane gauge is --we use latin indices $a,b,c,\cdots$ to denote the null-plane components of vectors ($a=+,1,2,-$)--:
\begin{align}\label{eq:2.4.11}
&A^a(x)=(2\pi)^{-3/2}\nonumber\\
&\times\sum_{\lambda=1,2}\int d\mu(\bm p)\varepsilon_\lambda(\bm p)^a\left(a_\lambda(\bm p)e^{-ipx}+a^\dagger_\lambda(\bm p)e^{ipx}\right),
\end{align}
in which only the physical degrees of freedom --the transversal polarization vectors-- appear. We are working with the following choice of polarization vectors \cite{AGPZ}:
\begin{equation*}
\varepsilon_1(\bm p)^a=\left(0;1;0;-\frac{p_1}{p_-}\right),\ \varepsilon_2(\bm p)^a=\left(0;0;1;-\frac{p_2}{p_-}\right),
\end{equation*}
\begin{equation}\label{eq:2.4.10.1}
\varepsilon_+(\bm p)^a=\left(1;-\frac{p_1}{p_-};-\frac{p_2}{p_-};\frac{p_\perp^2}{2p_-^2}\right),\ \varepsilon_-(\bm p)^a=\left(0;0;0;1\right).
\end{equation}
The photon emission and absorption field operators satisfy the following commutation rule:
\begin{equation}\label{eq:2.4.12}
\left[a_\lambda(\bm p);a_\sigma^\dagger(\bm q)\right]=2p_-\delta_{\lambda\sigma}\delta(\bm p-\bm q),
\end{equation}
from which the commutation relation for the gauge field can be derived:
\begin{equation}\label{eq:2.4.13}
\left[A^a(x);A^b(y)\right]=:iD^{ab}(x-y),
\end{equation}
with the commutation distribution:
\begin{align}\label{eq:2.4.14}
 D^{ab}(x)=&i(2\pi)^{-3}\int d^4p\text{sgn}\left(p_-\right)\delta(p^2)\nonumber\\
 &\times\left(g^{ab}-\frac{p^a\eta^b+\eta^ap^b}{p_-}\right)e^{-ipx}.
\end{align}
As we observe, the commutation distribution of the radiation field already contains nonlocal terms (gauge terms). This is different to what happens with the fermion field anticommutation distribution: It is covariant, and the nonlocal term appears only in its retarded part \cite{APQED1}. Therefore, since in the causal approach the causal distributions contain products of the positive- and negative-frequency parts of the (anti)commutation distributions of the quantized fields, the instantaneous term of the fermion field does not appear in loop calculations, a fact that was exploited in Yukawa's model \cite{AP2} and for the calculus of vacuum polarization in QED \cite{APQED1}. The situation here is different because the positive- and negative-frequency parts of the distribution in Eq. \eqref{eq:2.4.14} already contain nonlocal terms, so they will appear in the causal distributions corresponding to loop diagrams (in standard language). As a consequence, it is of fundamental importance for the solution of the equivalence problem to show that the ``gauge terms'' in Eq. \eqref{eq:2.4.14} do not contribute to any physical process at any order. It is clear that quantum gauge invariance is the key for that task to be accomplished, hence the present paper focuses on its implementation as a major part of the theory. The technique for constructing quantum gauge theories in instant dynamics CPT was developed by D{\"u}tsch, Hurth, Krahe and Scharf \cite{ScharfGauge1, ScharfGauge2, ScharfGauge3, ScharfGauge4}, then applied by D{\"u}tsch, Scharf and Aste \cite{ScharfYM1,ScharfYM2,ScharfYM3} to the construction of non-Abelian gauge theories, including the electroweak theory --see also Ref. \cite{ScharfGFT}--. Other important results regarding the uniqueness of the Yang-Mills theories can be found in Refs. \cite{Duetsch1} and \cite{Grigore1}.

In CPT, the distributional character of quantum fields is taken into account. As a consequence, the $S$-operator is a functional of the switching function $g\in\mathscr{S}(\mathbb{R}^4)$ \cite{Bogo} that multiply the coupling constant of the interaction, isolating the problem of infrared divergences; it is through the adiabatic limit $g\to 1$ that the real interaction is recovered. CPT is constructed over the axioms of translation invariance and causality, complemented with additional conditions as other symmetries and unitarity only at a later stage for the normalization of the solution. The scattering operator corresponding to an interaction regulated by $g\in\mathscr S(\mathbb{R}^4)$ is written as a formal series:
\begin{equation}\label{eq:3.2.3}
S(g)=1+\sum\limits_{n=1}^{+\infty}\frac{1}{n!}\int dXT_n(X)g(X);
\end{equation}
with $T_n(X)\equiv T_n(x_1;\ldots;x_n)$, $g(X)\equiv g(x_1)\ldots g(x_n)$, $dX\equiv d^4x_1\ldots d^4x_n$. This equation defines the transition distributions $T_n\in\mathscr S'(\mathbb{R}^{4n})$, which are symmetrical in the coordinates $x_1,\ldots,x_n$. The inverse operator $S(g)^{-1}$ is obtained as the formal inverse of $S(g)$:
\begin{equation}\label{eq:3.2.4}
S(g)^{-1}=1+\sum\limits_{n=1}^{+\infty}\frac{1}{n!}\int dX\widetilde T_n(X)g(X);
\end{equation}
\begin{equation*}
\widetilde T_n(X)=\sum\limits_{r=1}^n(-1)^r\sum\limits_{\begin{subarray}{l} X_1,\ldots,X_r\neq\emptyset\\ X_1\cup\ldots\cup X_r=X\\ X_j\cap X_k=\emptyset, \forall j\neq k \end{subarray}}T_{n_1}(X_1)\ldots T_{n_r}(X_r).
\end{equation*}

As a consequence of causality, the transition distributions are chronologically ordered --in the $x^+$ sense--:
\begin{align}\label{eq:3.2.17}
&T_n(X)=T_m(X_2)T_{n-m}(X_1)\ \text{for}\  X_1<X_2;\nonumber\\
&\left[T_n(X);T_m(Y)\right]=0\ \text{for}\  X\sim Y.
\end{align}
Because of this, we can define the advanced distribution of order $n$ as the following distribution:
\begin{equation}\label{eq:3.3.2}
A_n(Y;x_n)=\sum\limits_{\begin{subarray}{l} X\cup X'=Y\\ X\cap X'=\emptyset\end{subarray}}\widetilde T_m(X)T_{n-m}(X'\cup\left\{x_n\right\}),
\end{equation}
and the retarded distribution of order $n$ as:
\begin{equation}\label{eq:3.3.4}
R_n(Y;x_n)=\sum\limits_{\begin{subarray}{l} X\cup X'=Y\\ X\cap X'=\emptyset\end{subarray}}T_{n-m}(X'\cup\left\{x_n\right\})\widetilde T_m(X).
\end{equation}
In these distributions the $n$-point distribution appears once. Separating it from the other terms:
\begin{align}\label{eq:3.3.5}
&A_n(Y;x_n)=T_n(Y\cup\left\{x_n\right\})+A'_n(Y;x_n),\nonumber\\
&R_n(Y;x_n)=T_n(Y\cup\left\{x_n\right\})+R'_n(Y;x_n),
\end{align}
with $A'_n$ and $R'_n$ the advanced subsidiary distribution and the retarded subsidiary distribution, respectively, which do not contain $T_n$. The transition distribution of order $n$ is then equal to:
\begin{align}\label{eq:3.3.8}
T_n(Y\cup\left\{x_n\right\})&=R_n(Y;x_n)-R'_n(Y;x_n).
\end{align}
Therefore, the $n$-point distribution can be found by obtaining the retarded distribution of order $n$, which can be done by splitting \cite{Division1,Division2,Division3} the causal distribution of order $n$:
\begin{align}\label{eq:3.3.9}
D_n(Y;x_n)&:=R_n(Y;x_n)-A_n(Y;x_n)\nonumber\\
&=R'_n(Y;x_n)-A'_n(Y;x_n),
\end{align}
which, at the light of the last equality, can be constructed with the knowledge of the transition distributions up to order $n-1$, only. It must be done as follows: The causal distribution has, in general, the following form:
\begin{equation}\label{eq:3.5.1}
D_n(x_1;\ldots;x_n)=\sum_kd^k_n(x_1;\ldots;x_n)\normord{C_k(u^A)},
\end{equation}
with $d^k_n$ a numerical distribution and $\normord{C_k(u^A)}$ a Wick's monomial of the quantized free field operators $u^A$. The support properties of the operator-valued distribution are encoded into the numerical distribution $d^k_n$, hence it is sufficient to split it. Using translation invariance, define the numerical distribution $d\in\mathscr{S}(\mathbb{R}^{4n-4})$ as:
\begin{equation}\label{eq:3.5.6}
d(x):=d^k_n(x_1-x_n;\ldots;x_{n-1}-x_n;0),
\end{equation}
with $\text{supp}(d)\subseteq\Gamma^+_{n-1}(0)\cup\Gamma^-_{n-1}(0)$. It must be split as:
\begin{equation}\label{eq:3.5.7}
d=r-a;\ \text{supp}(r)\subseteq\Gamma^+_{n-1}(0),\ \text{supp}(a)\subseteq\Gamma^-_{n-1}(0).
\end{equation}
Here we are denoting:
\begin{align*}
&\Gamma^+_n(0):=\Big\{(x_1;\cdots;x_n)\in\mathbb{M}^n\ \Big|\ \forall j\in\left\{1,\cdots, n\right\}:\nonumber\\
&x_j^+\geq 0\wedge\ \left( \exists x_k\in\overline{V^+}(0) (k\neq j) : x_j\in\widetilde V^+(x_k)\right)\Big\} ,
\end{align*}
with $V^\pm(x)$ the interior of the future or past, respectively, light-cone with vertex at the point $x$, $\overline{V^\pm}(x)$ its closure and $\widetilde V^\pm(x)$ the union of its closure and the $x^-$ axis. An analogous definition holds for $\Gamma^-_n(0)$. We are using Schwartz's multi-index notation \cite{SchwartzM}.

To perform the splitting, it is crucial to remember that the product of a distribution by a discontinuous function can be ill-defined if the distribution has a singularity precisely on the discontinuity surface of the function. In our case we then need to control the behavior of the causal distribution near the splitting region, which is the $x^-$-axis. This can be done by applying the concept of quasiasymptotics by selected variable \cite{Vladimirov}:

\textit{Definition.---} Let $d\in\mathscr{S}'(\mathbb{R}^m)$ be a distribution, and let $\rho$ be a continuous positive function. If the (distributional) limit
\begin{equation}\label{eq:3.5.9}
\lim_{s\to 0^+}\rho(s)s^{3m/4}d\left(sx^+;sx^\perp;x^-\right)=d_-(x)
\end{equation}
exists in $\mathscr{S}'(\mathbb{R}^m)$ and is non-null, then the distribution $d_-$ is called the quasiasymptotics of $d$ at the $x^-$ axis, with regard to the function $\rho$.

One can then show \cite{Vladimirov, ScharfFQED} that for every $a>0$: $\displaystyle{\lim_{s\to0^+}\rho(as)/\rho(s)=a^\alpha}$ for some $\alpha\in\mathbb{R}$. This number, denoted by $\omega_-$, characterizes the distribution, and is called its singular order at the $x^-$ axis.

In momentum space the following splitting formulas are found: For negative singular order, $\omega_-<0$:
\begin{align}
\hat r(p)=\frac{i}{2\pi}\int\limits_{-\infty}^{+\infty} \frac{\hat d(p_+-k;\bm p)}{k+i0^+}dk.\label{eq:3.6.9}
\end{align}
For non-negative singular order, $\omega_-\geq0$, the retarded distribution normalized at $\left(q_+;q_\perp;p_-\right)$ is:
\begin{align}\label{eq:3.6.47}
&\hat r_q(p)=\frac{i}{2\pi}\int\limits_{-\infty}^{+\infty}\frac{dk}{k+i0^+}\Bigg\{\hat d(p_+-k;\bm p)\nonumber\\
&-\sum\limits_{|c|=0}^{\lfloor\omega_-\rfloor}\frac{1}{c!}(p_{+,\alpha}-q_{+,\alpha})^cD^c_{+,\alpha}\hat d(q_+-k;q_\perp;p_-)\Bigg\}.
\end{align}
Particularly, the central solution is the one normalized at the line $(0;0_\perp;p_-)$.

Finally, if $r_1$ and $r_2$ are two solutions of the splitting problem, then they could differ by normalization terms supported at the $x^-$ axis. In momentum space:
\begin{equation}\label{eq:3.7.5}
\hat r_1(p)-\hat r_2(p)=\sum\limits_{|b|=0}^{M}\widehat C_b\left(p_-\right)p_{+,\perp}^b,
\end{equation}
with $\widehat C_b\left(p_-\right)$ some distributions of the variable $p_-$. The procedure of fixing them by the imposition of physical requirements is called the normalization process.

This paper has the following structure. The construction of the gauge invariant theory is performed in Sec. \ref{sec:qgi}, while its more direct consequences for QED, Ward-Takahashi's identities, are presented in Sec. \ref{sec:WTI}. With those tools, in Sec. \ref{sec:self} we show the calculation of the electron self-energy and its normalization. Sec. \ref{sec:Conc} contains our conclusions and perspectives of future work. In appendix, finally, we prove the normalization of the physical $S$-matrix of null-plane QED.

\section{Quantum gauge invariance}\label{sec:qgi}

As we have seen in Ref. \cite{APQED1}, the gauge terms of the radiation field commutation distribution do not contribute to M{\o}ller's scattering, which is an expression of gauge invariance. It is our aim in this section to prove that they in fact do not contribute to any process, in other words, that the scattering operator can be constructed with the covariant terms only. To do that, we will get back to the quantization procedure of the massless vector field --see Ap. A in Ref. \cite{APS22}--. Such a procedure had the meaning of constructing Fock's space and the operators as well as operator-valued distributions acting on it. This Fock's space consists on physical states, only, whose wavefunctions are the positive-frequency part of the classical field solutions. As a consequence, only physical polarization states are quantized, leading to the instantaneous term in the commutation distribution of the field operator. If we want to use a covariant commutation distribution, it will be necessary to quantize the non-physical polarization states as well, and, accordingly, to extend Fock's space in order to contain the non-physical states created by their corresponding field operators. This extension must be done in such a way that the physical content of the theory is not altered, as will be shown in the following paragraphs.

\subsection{Fock's space extension}

Vector massless fields have only two degrees of freedom, identified with the transversal polarizations $\varepsilon_\alpha(x)^a$ ($\alpha=1,2$). Fock's space of these physical states will be called the physical subspace, $\mathcal F_{\text{phys}}$, of the complete Fock's space, $\mathcal F$, because the physical potentials will be those which satisfy both the null-plane and Lorenz's gauge conditions. All the sectors of $\mathcal F$ different from the physical subspace are inaccessible to experimental observation; therefore, there are no reasons to expect that the expression of the quantized field operator has the same form as its classical version: Fock's space extension is an eminently mathematical process.

If the gauge conditions are not imposed to the quantized field operator, then we must quantize all the four polarization states $\lambda=+,1,2,-$, hence we introduce not two, but four sets of emission and absorption operators $a_\lambda^\dagger(f)$ and $a_\lambda(f)$. In order to have a positive-definite inner product in the complete Fock's space, we impose that they must satisfy:
\begin{equation}\label{7.1.1}
\left[a_\lambda(\bm p);a_\lambda^\dagger(\bm q)\right]=2p_-\delta(\bm p-\bm q).
\end{equation}
As a consequence, all states, including the non-transverse ones, have positive-definite norm. The most natural extension of the quantized radiation field operator would be:
\begin{align}\label{eq:7.1.2}
&A^a(x)=(2\pi)^{-3/2}\nonumber\\
&\times\sum_\lambda\int d\mu(\bm p)\varepsilon_\lambda(\bm p)^a\left(a_\lambda(\bm p)e^{-ipx}+a_\lambda^\dagger(\bm p)e^{ipx}\right),
\end{align}
with the sum extended to all the polarizations, including the non-physical ones $\lambda=+,-$. However, such a field operator has the following commutation distribution:
\begin{align*}
&\left[A^a(x);A^b(y)\right]=(2\pi)^{-3}\int d^4p\delta(p^2)\Theta(p_-)\nonumber\\
&\quad\times\left(e^{-ip(x-y)}-e^{ip(x-y)}\right)\Big(-g^{ab}+\frac{p^a\eta^b+\eta^ap^b}{p_-}\nonumber\\
&\quad+\varepsilon_+(\bm p)^a\varepsilon_+(\bm p)^b+\varepsilon_-(\bm p)^a\varepsilon_-(\bm p)^b\Big),
\end{align*}
which also exhibits instantaneous terms. In order to obtain a covariant commutation distribution, as we need, we must only have the term $-g^{ab}$ inside the parentheses of the second line of the above equation. We will see that this is possible with a convenient redefinition of the quantized field operators associated to non-physical polarizations. Let us write:
\begin{equation}\label{eq:7.1.3}
A^a(x)=(2\pi)^{-3/2}\sum_\lambda\int d\mu(\bm p)\varepsilon_\lambda(\bm p)^aA_\lambda(\bm p;x).
\end{equation}
This expression coincides with the real quantized field operator for the physical polarization states if:
\begin{equation}\label{eq:7.1.4}
A_{1,2}(\bm p;x)=a_{1,2}(\bm p)e^{-ipx}+a_{1,2}^\dagger(\bm p)e^{ipx}.
\end{equation}
The commutator is then:
\begin{align}\label{eq:7.1.5}
&\left[A^a(x);A^b(y)\right]=(2\pi)^{-3}\int d\mu(\bm p)d\mu(\bm q)\nonumber\\
&\quad\times\sum_{\lambda,\lambda'}\varepsilon_\lambda(\bm p)^a\varepsilon_{\lambda'}(\bm q)^b\left[A_\lambda(\bm p;x);A_{\lambda'}(\bm q;y)\right].
\end{align}
Taking in mind the completeness relation of the polarization vectors, which is:
\begin{equation}\label{eq:7.1.6}
\sum_{\lambda,\lambda'}g_{\lambda\lambda'}\varepsilon_\lambda(\bm p)_a\varepsilon_{\lambda'}(\bm p)_b=g_{ab},
\end{equation}
the commutation distribution will be the one we need if:
\begin{align}\label{eq:7.1.7}
&\left[A_\lambda(\bm p;x);A_{\lambda'}(\bm q;y)\right]=\nonumber\\
&\quad-2p_-\delta(\bm p-\bm q)g_{\lambda\lambda'}\left(e^{-i(px-qy)}-e^{i(px-qy)}\right).
\end{align}
This relation is satisfied for $\lambda=1,2$ with $A_{1,2}(\bm p;x)$ from Eq. \eqref{eq:7.1.4}. The other components are also obtained if we take:
\begin{equation}\label{eq:7.1.8}
A_\pm(\bm p;x)=a_\pm(\bm p)e^{-ipx}-a_\mp^\dagger(\bm p)e^{ipx}.
\end{equation}
Accordingly, we shall define the quantized massless vector field operator as:
\begin{align}\label{eq:7.1.9}
&A^a(x):=(2\pi)^{-3/2}\sum_\lambda\int d\mu(\bm p)\varepsilon_\lambda(\bm p)^a\nonumber\\
&\times\left(a_\lambda(\bm p)e^{-ipx}-\sum_{\tau}g_{\lambda\tau}a_\tau^\dagger(\bm p)e^{ipx}\right),
\end{align}
which satisfies a covariant commutation relation:
\begin{equation}\label{eq:7.1.10}
\left[A^a(x);A^b(y)\right]=ig^{ab}D_0(x-y),
\end{equation}
with $D_0(x)$ the massless Jordan-Pauli's distribution. It will also be useful to have the commutator of the negative and positive frequency parts of the field \footnote{The sub-index indicating the frequency sign will be put at the right of the index of the component. For example, the positive frequency part of the longitudinal component of the field is: $A\indices{^-_+}(x)=A_{++}(x)$, and so on.}:
\begin{align}
&\left[A\indices{^a_-}(x);A\indices{^b_+}(y)\right]=-(2\pi)^{-3}\sum_{\lambda,\lambda'}\int d\mu(\bm p)d\mu(\bm q)\nonumber\\
&\quad\times\varepsilon_\lambda(\bm p)^a\varepsilon_{\lambda'}(\bm q)^b\sum_{\tau'}g_{\lambda'\tau'}\left[a_\lambda(\bm p);a_{\tau'}^\dagger(\bm q)\right]e^{-i(px-qy)}\nonumber\\
&=-(2\pi)^{-3}\int d\mu(\bm p)\sum_{\lambda\lambda'}g_{\lambda\lambda'}\varepsilon_\lambda(\bm p)^a\varepsilon_{\lambda'}(\bm p)^be^{-ip(x-y)}\nonumber\\
&=ig^{ab}D_+(x-y).\label{eq:7.1.11}
\end{align}
And, analogously:
\begin{equation}\label{eq:7.1.12}
\left[A\indices{^a_+}(x);A\indices{^b_-}(y)\right]=ig^{ab}D_-(x-y).
\end{equation}
This is the desired result. Now Fock's space contains photons with the four polarization degrees, among which only the transversal ones can constitute asymptotically free states. In this extended space the positive-definite Hamiltonian operator is:
\begin{equation}\label{eq:7.1.13}
\bm{\mathsf P}_+=\int d\mu(\bm p)p_+\sum_\lambda a_\lambda^\dagger(\bm p)a_\lambda(\bm p),
\end{equation}
because with it the field operator of Eq. \eqref{eq:7.1.9} satisfies Heisenberg's equation of motion:
\begin{align}
i\partial_+ A^a(x)&=(2\pi)^{-3}\sum_\lambda\int d\mu(\bm p)p_+\varepsilon_\lambda(\bm p)^a\nonumber\\
&\quad\times\left(a_\lambda(\bm p)e^{-ipx}+\sum_{\tau}g_{\lambda\tau}a_\tau^\dagger(\bm p)e^{ipx}\right)\nonumber\\
&=\left[A^a(x);\bm{\mathsf P}_+\right].\label{eq:7.1.14}
\end{align}
What remains in order to prove that the gauge terms in the radiation field commutation distribution in Eq. \eqref{eq:2.4.14} do not contribute is to show that the physics does not change when the extended field is used. The proof of this statement is the topic of the following paragraphs of this section.

\subsection{Poincar{\'e}'s invariance of the physical subspace}

Let us consider the null-plane and Lorenz's gauge conditions when applied to the quantized field operator. Since the ``$+$'' component of all polarization vectors, except the one of $\varepsilon_+(\bm p)^a$, is null [see Eq. \eqref{eq:2.4.10.1}], we have that:
\begin{align}\label{eq:7.1.15}
&A^+(x)=(2\pi)^{-3/2}\nonumber\\
&\times\int d\mu(\bm p)\varepsilon_+(\bm p)^+\left(a_+(\bm p)e^{-ipx}-a_-^\dagger(\bm p)e^{ipx}\right).
\end{align}
Also, the divergence of the quantized field operator is:
\begin{align}\label{eq:7.1.16}
&\partial_a A^a(x)=-i(2\pi)^{-3/2}\sum_\lambda\int d\mu(\bm p)p_a\varepsilon_\lambda(\bm p)^a\nonumber\\
&\times\left(a_\lambda(\bm p)e^{-ipx}+\sum_\tau g_{\lambda\tau}a_\tau^\dagger(\bm p)e^{ipx}\right).
\end{align}
But, as it can be seen in Eq. \eqref{eq:2.4.10.1}, the polarization vectors are chosen in such a way that:
\begin{equation}\label{eq:7.1.17}
\varepsilon_{1,2}(\bm p)_-=0\ ,\  \varepsilon_{1,2}(\bm p)_+=-\frac{p_\alpha\varepsilon_{1,2}(\bm p)^\alpha}{p_-},
\end{equation}
which imply that they are orthogonal to the momentum $p^a$: $p_a\varepsilon_{1,2}(\bm p)^a=0$. Therefore, in the sum in Eq. \eqref{eq:7.1.16} only the polarization states $\lambda=+,-$ contribute:
\begin{align}\label{eq:7.1.18}
&\partial_a A^a(x)=-i(2\pi)^{-3/2}\sum_{\lambda=+,-}\int d\mu(\bm p)p_a\varepsilon_\lambda(\bm p)^a\nonumber\\
&\times\left(a_\lambda(\bm p)e^{-ipx}+\sum_\tau g_{\lambda\tau}a_\tau^\dagger(\bm p)e^{ipx}\right),
\end{align}
and only the emission and absorption field operators $a_\pm^\dagger(\bm p)$ and $a_\pm(\bm p)$ appear. Since every state in $\mathcal F_{\text{phys}}$ has polarizations $\lambda=1,2$, Eqs. \eqref{eq:7.1.15} and \eqref{eq:7.1.18} mean that this subspace can be characterized by the accomplishment of the gauge conditions as matrix elements:
\begin{equation}\label{eq:7.1.19}
\forall \Phi,\Psi\in\mathcal F_{\text{phys}}:\ \left(\Phi;A^+(x)\Psi\right)=0\ \wedge\  \left(\Phi;\partial_a A^a(x)\Psi\right)=0.
\end{equation}

Let $f=(f_a)$ be a wave-function in the extended one-particle Hilbert's space; it is transformed under a Poincar{\'e}'s transformation by the classical field law:
\begin{equation}\label{eq:7.1.20}
\left(U(a;\Lambda)f\right)(x)=\Lambda^{-1}f\left(\Lambda^{-1}(x-a)\right).
\end{equation}
Over the space of these functions we define the operator $A(f)$ according to:
\begin{equation}\label{eq:7.1.21}
A(f):=\int d^4xf_a(x)A^a(x);
\end{equation}
this is the smearing of the operator-valued distribution $A^a(x)$ over the test-function $f_a$. The operator of Poincar{\'e}'s transformation which acts on the extended Fock's space is the operator $\bm{\mathsf U}(a;\Lambda)$ such that:
\begin{equation}\label{eq:7.1.22}
\bm{\mathsf U}(a;\Lambda)A(f)\bm{\mathsf U}(a;\Lambda)^{-1}=A\left(U(a;\Lambda)f\right).
\end{equation}
Hence, using Eqs. \eqref{eq:7.1.20} and \eqref{eq:7.1.21} we obtain the transformation law of the quantized field operator:
\begin{equation}\label{eq:7.1.23}
\bm{\mathsf U}(a;\Lambda)A^a(x)\bm{\mathsf U}(a;\Lambda)^{-1}=\left(\Lambda^{-1}\right)\indices{^a_b}A^b(\Lambda x+a).
\end{equation}
The application of the ``$\dagger$'' adjoint to this equation allows us to see that the operator $\bm{\mathsf U}(a;\Lambda)$ is not unitary because $A^a(x)$ is not Hermitian. This does not violate Wigner's theorem: Poincar{\'e}'s transformations are symmetries in the real world, so that they must be symmetries in the physical subspace only. In order to show that this is the case, denote by $\mathcal F_L$ the subspace of $\mathcal F$ which satisfies Lorenz's gauge condition. In Ref. \cite{Rohrlich2} it is shown that this gauge condition is compatible with the null-plane one in the free case --which is always the case in CPT--, and, equally important, that these two conditions determine the gauge completely --\textit{i.e.}, there is no remnant gauge symmetry after the imposition of them--; therefore, we can affirm:
\begin{align}\label{eq:7.1.24}
\forall \Phi'\in&\mathcal F_L:\ \exists!\Phi\in\mathcal F_{\text{phys}}:\nonumber\\
& \Phi'_a(x)=\Phi_a(x)+\partial_a\Lambda(x),\ \square\Lambda(x)=0.
\end{align}
We denote the projection onto the physical subspace by: $\Phi=P\Phi'$. Now, the inner product in the one-particle space is:
\begin{align}\label{eq:7.1.26}
(f;g)_1&=\sum_a i\int f_a(x)^*\overleftrightarrow\partial_-g_a(x)d^3\bm x\nonumber\\
&=\sum_a\int d\mu(\bm p)\hat f_a(\bm p)^*\hat g_a(\bm p).
\end{align}
Let $\Phi,\Psi\in\mathcal F_{\text{phys}}$ be one-particle states, with wave-functions $\Phi_a$ and $\Psi_a$, respectively. They satisfy the null-plane gauge: $\Phi^+=0=\Psi^+$, hence their inner product can be put in Poincar{\'e}'s invariant form:
\begin{align}\label{eq:7.1.27}
(\Phi;\Psi)&=\sum_\alpha i\int\Phi_\alpha(x)^*\overleftrightarrow\partial_-\Psi_\alpha(x)d^3\bm x\nonumber\\
&=-\sum_\alpha i\int\Phi_\alpha(x)^*\overleftrightarrow\partial_-\Psi^\alpha(x)d^3\bm x\nonumber\\
&=-(\Phi_a;\Psi^a).
\end{align}
Let us apply now a reference frame transformation. In the general case the value of $\Phi^+=0$ changes, and the transformed states $\Phi'$ and $\Psi'$ are no more in $\mathcal F_{\text{phys}}$, but they are in $\mathcal F_L$ yet. Applying the operator $P$ to obtain again states in $\mathcal F_{\text{phys}}$:
\begin{equation}\label{eq:7.1.28}
\widetilde\Phi=P\bm{\mathsf U}(a;\Lambda)\Phi=:\widetilde{\bm{\mathsf U}}(a;\Lambda)\Phi,\ \widetilde\Psi=\widetilde{\bm{\mathsf U}}(a;\Lambda)\Psi.
\end{equation}
Therefore, the operator which transforms the states in $\mathcal F_{\text{phys}}$ is not $\bm{\mathsf U}(a;\Lambda)$, but $\widetilde{\bm{\mathsf U}}(a;\Lambda)$, and it is this one that must be unitary on $\mathcal F_{\text{phys}}$. Effectively it is: Since $\widetilde \Phi$ and $\Phi'$ are related by a gauge transformation, as well as $\widetilde\Psi$ and $\Psi'$, we write:
\begin{equation}\label{eq:7.1.29}
\Phi'_a=\widetilde\Phi_a+\partial_a\chi,\ \Psi'_a=\widetilde\Psi_a+\partial_a\Lambda.
\end{equation}
Then the inner product in Eq. \eqref{eq:7.1.27}, which is Poincar{\'e}'s invariant, is equal to:
\begin{align}\label{eq:7.1.30}
&(\Phi;\Psi)=-(\Phi'_a;\Psi'^a)\nonumber\\
&=\left(\widetilde\Phi;\widetilde\Psi\right)-\left(\partial_a\chi;\widetilde\Psi^a\right)-\left(\widetilde\Phi_a;\partial^a\Lambda\right)-\left(\partial_a\chi;\partial^a\Lambda\right).
\end{align}
It is easy to see that the second and third terms in the last equality are null, because $\widetilde\Phi$ and $\widetilde\Psi$ satisfy the gauge conditions and because they are assumed to vanish at infinity (asymptotic conditions). Using again the asymptotic conditions, the last term takes the form:
\begin{align*}
&(\partial_a\chi;\partial^a\Lambda)=(\partial_+\chi;\partial_-\Lambda)+(\partial_-\chi;\partial_+\Lambda)-(\partial_\alpha\chi;\partial_\alpha\Lambda)\\
&=-(\partial_+\partial_-\chi;\Lambda)-(\chi;\partial_+\partial_-\Lambda)+\frac{1}{2}(\partial_\perp^2\chi;\Lambda)+\frac{1}{2}(\chi;\partial_\perp^2\Lambda)\\
&=-\frac{1}{2}(\square\chi;\Lambda)-\frac{1}{2}(\chi;\square\Lambda),
\end{align*}
which are null as established in Eq. \eqref{eq:7.1.24}. In conclusion:
\begin{equation}\label{eq:7.1.31}
(\Phi;\Psi)=\left(\widetilde\Phi;\widetilde\Psi\right)=\left(\widetilde{\bm{\mathsf U}}(a;\Lambda)\Phi;\widetilde{\bm{\mathsf U}}(a;\Lambda)\Psi\right),
\end{equation}
and Poincar{\'e}'s transformations operator in $\mathcal F_{\text{phys}}$, $\widetilde{\bm{\mathsf U}}(a;\Lambda)$, is a unitary operator. This shows that the physical subspace is the same in all reference frames.

\subsection{Quantum gauge transformations}

As we have said, the extension of Fock's space is not a physical procedure, but a mathematical one. Accordingly, we now have to impose that the physical content of the theory is independent of this extension.

\textit{Definition.---} A quantum gauge transformation is a transformation of the quantized field, or equivalently, of the quantized field operator, which depends continuously on a constant parameter $\lambda$ and has the form
\begin{equation}\label{eq:7.1.35}
A'(f)=e^{-i\lambda Q}A(f)e^{i\lambda Q}\ \Leftrightarrow\  A'^a(x)=e^{-i\lambda Q}A^a(x)e^{i\lambda Q},
\end{equation}
form which warrants (1) that the commutation distribution of the transformed field operator does not change, and (2) that the quantized field operator still satisfies the equation of motion $\square A'^a(x)=0$. Also, we impose the condition that $e^{i\lambda Q}$ is an operator which leaves invariant the states of $\mathcal F_{\text{phys}}$. The operator $Q$, generator of the quantum gauge transformations, is called the gauge charge operator.

The most immediate consequence of this definition is that the gauge charge operator annihilates physical states, hence:
\begin{equation}\label{eq:7.1.36}
\mathcal F_{\text{phys}}\subseteq \text{Ker}(Q).
\end{equation}
But we do not know yet under what conditions the equality holds in Eq. \eqref{eq:7.1.36}. Let us start by establishing:

\textbf{Lemma:} \textit{The gauge charge operator $Q$, as well as its adjoint $Q^\dagger$, is constructed with non-physical emission and absorption operators, only.} 

\textit{Proof:} In first place, let us see that the non-physical states are orthogonal to the physical ones; in effect, let $a_{\text{phys}}^\dagger$ and $a_{\text{nph}}^\dagger$ be emission operators of physical and non-physical particles, respectively. Since they commute:
\begin{align}\label{eq:7.1.37}
&\left(a_{\text{phys}}^\dagger\Omega;a_{\text{nph}}^\dagger\Omega\right)=\left(\Omega;a_{\text{phys}}a_{\text{nph}}^\dagger\Omega\right)\nonumber\\
&\quad=\left(\Omega;a_{\text{nph}}^\dagger a_{\text{phys}}\Omega\right)=0.
\end{align}
Consider now two states $\Phi,\Psi\in\mathcal F_{\text{phys}}$:
\begin{equation}\label{eq:7.1.38}
0=(\Phi;Q\Psi)=\left(Q^\dagger\Phi;\Psi\right)\ \Rightarrow\  Q^\dagger\Phi\in\mathcal{F}_{\text{phys}}^\perp.
\end{equation}
Now, every operator acting on Fock's space can be written as a function of emission and absorption operators. Particularly, in order to $Q$ to annihilate every physical state it is mandatory that all the terms in it have an absorption operator of a non-physical particle at the right. Let us suppose that one of the terms has a physical emission operator at the left:
\begin{equation}\label{eq:7.1.39}
Q\sim a_{\text{phys}}^\dagger a_{\text{nph}}+\ldots
\end{equation}
Then, applying the adjoint to a physical state $\Phi$:
\begin{equation}\label{eq:7.1.40}
Q^\dagger\Phi\sim a_{\text{nph}}^\dagger a_{\text{phys}}\Phi+\ldots,
\end{equation}
which in general still contains physical particles, in contradiction with Eq. \eqref{eq:7.1.38}. Hence, $Q$ could not contain physical emission operators, and the same is clearly true for physical absorption operators. From this, it follows that $Q^\dagger$ also contains non-physical emission and absorption operators, hence $\forall\Phi\in\mathcal F_{\text{phys}}$: $Q^\dagger\Phi=0$. $\blacksquare$

As a consequence of this lemma, Eq. \eqref{eq:7.1.36} must be substituted by:
\begin{equation}\label{eq:7.1.41}
\mathcal F_{\text{phys}}\subseteq\text{Ker}(Q)\cap\text{Ker}\left(Q^\dagger\right)=\text{Ker}\left(\left\{Q;Q^\dagger\right\}\right).
\end{equation}

Expanding the exponentials in Eq. \eqref{eq:7.1.35} as a series in the parameter $\lambda$, we will find that:
\begin{align}\label{eq:7.1.43}
A'^a(x)=&A^a(x)-i\lambda\left[Q;A^a(x)\right]\nonumber\\
&-\frac{\lambda^2}{2}\left[Q;[Q; A^a(x)]\right]+\mathscr{O}(\lambda^3).
\end{align}
By virtue of the lemma, all the commutators in Eq. \eqref{eq:7.1.43} are non-null only for the non-physical part of the quantized field operator $A^a(x)$; in other words, the quantum gauge transformation does not modify the dynamical (physical) part of the quantized field, as required, and all the quantities constructed with $A^a(x)$ will have the same matrix elements on the physical subspace: This is quantum gauge invariance.

Now, every operator $Q$, constructed with non-physical emission and absorption operators in such a way that an emission operator is at the left and an absorption operator is at the right of every term, can be used as a generator of a quantum gauge transformation. However, such a general gauge charge could originate a transformed gauge field which is a composite operator. We must impose that $Q$ is a quadratic operator in order to maintain $A^a(x)$ as a simple field operator.

For a general gauge charge, Eq. \eqref{eq:7.1.41} is satisfied with the symbol ``$\subset$''; the equality holds if $\left\{Q;Q^\dagger\right\}$ is an operator in which all the non-physical absorption operators appear. In order to obtain such a gauge charge, note that $[Q;A^a(x)]$ is always a linear combination of non-physical emission and absorption operators. Therefore, regarding Eq. \eqref{eq:7.1.43}, define the field operator $u(x)$ by:
\begin{equation}\label{eq:7.1.45}
i\partial_a u(x):=\left[Q;A_a(x)\right].
\end{equation}
Eq. \eqref{eq:7.1.43} adopts the form:
\begin{equation}\label{eq:7.1.46}
A'^a(x)=A^a(x)+\lambda\partial^au(x)-\frac{i\lambda^2}{2}\left[Q;\partial^au(x)\right]+\mathscr{O}(\lambda^3).
\end{equation}
But, by definition, $A'^a(x)$ must satisfy Klein-Gordon-Fock's equation, which is possible if $u(x)$ satisfies:
\begin{equation}\label{eq:7.1.47}
\square u(x)=0.
\end{equation}
The solution for this equation is:
\begin{equation}\label{eq:7.1.48}
u(x)=\int\limits_{y^+=y_0^+}u(y)\overleftrightarrow\partial_-^yD_0(y-x).
\end{equation}
Deriving Eq. \eqref{eq:7.1.48} and recognizing the commutation distribution of the extended gauge field operator of Eq. \eqref{eq:7.1.10}:
\begin{equation}\label{eq:7.1.50}
i\partial_au(x)=\left[\int\limits_{y^+=y_0^+}\partial_bA^b(y)\overleftrightarrow\partial_- u(y)d^3\bm y;A_a(x)\right],
\end{equation}
from which we identify, by comparison with Eq. \eqref{eq:7.1.45}, that the gauge charge operator is:
\begin{equation}\label{eq:7.1.51}
Q=\int\limits_{x^+=x_0^+}\partial_aA^a(x)\overleftrightarrow\partial_- u(x)d^3\bm x.
\end{equation}

On the other hand, Eq. \eqref{eq:7.1.47} implies that the field operator $u(x)$ has the form:
\begin{equation}\label{eq:7.1.52}
u(x)=(2\pi)^{-3/2}\int d\mu(\bm p)\left(c_2(\bm p)e^{-ipx}+c_1^\dagger(\bm p)e^{ipx}\right).
\end{equation}
The field operators $c_1(\bm p)$ and $c_2(\bm p)$ are, at the moment, unknown, but they belong to non-physical ``particles''\footnote{This is, of course, an abuse of language, because the concept of particle can only be defined in the asymptotically free region, whereas a ``non-physical particle'' cannot reach that region.}. Joining Eqs. \eqref{eq:7.1.18} and \eqref{eq:7.1.52} into Eq. \eqref{eq:7.1.51} and by an elemental integration, we find:
\begin{align}\label{eq:7.1.53}
Q=&\sum\limits_{\lambda=+,-}\int d\mu(\bm p)p_a\varepsilon_\lambda(\bm p)^a\nonumber\\
&\times\left(a_\lambda(\bm p)c_1^\dagger(\bm p)-\sum_\tau g_{\lambda\tau}a_\tau^\dagger(\bm p)c_2(\bm p)\right).
\end{align}
By construction, this is the most general form that a quadratic gauge charge operator can have. This formula reveals: Since $Q$ must have absorption operators to the right, and this does not occur in the first term of Eq. \eqref{eq:7.1.53}, we must commute $a_\lambda(\bm p)$ with $c_1^\dagger(\bm p)$; but, in doing that, if $c_1(\bm p)$ were one of $a_+(\bm p)$ or $a_-(\bm p)$, then $Q$ would gain a constant term and does not annihilate the physical states. This is impossible. In consequence, it is necessary to extend Fock's space even more in order to contain field operators $c_1(\bm p)$ and $c_2(\bm p)$ and their adjoints, associated to non-physical particles and which are different from $a_+(\bm p)$ and $a_-(\bm p)$. The quantized field operator $u(x)$, as a new non-physical field, is called ghost field. As we see, the ghost field is indispensable for quantum gauge invariance. Since $c_1^\dagger(\bm p)$ corresponds to a new particle, it can be simply commutated with $a_\lambda(\bm p)$. From this in Eq. \eqref{eq:7.1.53} it follows that:
\begin{align}\label{eq:7.1.54}
Q=&\sum\limits_{\lambda=+,-}\int d\mu(\bm p)p_a\varepsilon_\lambda(\bm p)^a\nonumber\\
&\times\left(c_1^\dagger(\bm p)a_\lambda(\bm p)-\sum_\tau g_{\lambda\tau}a_\tau^\dagger(\bm p)c_2(\bm p)\right).
\end{align}
The explicit form of the polarization vectors of the massless vector field are given in Eq. \eqref{eq:2.4.10.1}, from which it follows that:
\begin{equation}\label{eq:7.1.59}
p_a\varepsilon_+(\bm p)^a=p_+-\frac{p_\perp^2}{2p_-}=\frac{p^2}{2p_-},\  p_a\varepsilon_-(\bm p)^a=p_-.
\end{equation}
Introducing this into Eq. \eqref{eq:7.1.54} and taking in mind that for the massless field it is $p^2=0$, we arrive at the final expression:
\begin{equation}\label{eq:7.1.60}
Q=\int d\mu(\bm p)p_-\left(c_1^\dagger(\bm p)a_-(\bm p)-a_+^\dagger(\bm p)c_2(\bm p)\right).
\end{equation}
The adjoint operator is:
\begin{equation}\label{eq:7.1.61}
Q^\dagger=\int d\mu(\bm p)p_-\left(a_-^\dagger(\bm p)c_1(\bm p)-c_2^\dagger(\bm p)a_+(\bm p)\right).
\end{equation}

In order to characterize the physical subspace we are interested in the anticommutator of $Q$ and $Q^\dagger$ [see Eq. \eqref{eq:7.1.41}]. Using Eqs. \eqref{eq:7.1.60} and \eqref{eq:7.1.61}:
\begin{align}
&\left\{Q;Q^\dagger\right\}=\int d\mu(\bm p)d\mu(\bm q)p_-q_-\Big(2p_-\delta(\bm p-\bm q)c_1^\dagger(\bm p)c_1(\bm p)\nonumber\\
&+2p_-\delta(\bm p-\bm q)c_2^\dagger(\bm p)c_2(\bm p)+\left\{c_1(\bm q);c_1^\dagger(\bm p)\right\}a_-^\dagger(\bm q)a_-(\bm p)\nonumber\\
&+\left\{c_2(\bm p);c_2^\dagger(\bm q)\right\}a_+^\dagger(\bm p)a_+(\bm q)\Big).\label{eq:7.1.62}
\end{align}
The first two terms in this equation have the desired form. For the other two, if $c_1(\bm p)$ and $c_2(\bm p)$ were boson operators, then the anticommutator $\left\{Q;Q^\dagger\right\}$ would have terms of the form $a_-^\dagger(\bm q)c_1^\dagger(\bm p)c_1(\bm q)a_-(\bm p)$, and there will be non-physical states annihilated by $\left\{Q;Q^\dagger\right\}$, for example, a one particle state with longitudinal polarization. As this is not required and contradicts our hypothesis on the physical degrees of freedom of the gauge field, the quantized field operator $u$ must be a fermion field, with emission and absorption field operators subjected to the anticommutation rules:
\begin{align}\label{eq:7.1.63}
&\left\{c_1(\bm p);c_1^\dagger(\bm q)\right\}=2p_-\delta(\bm p-\bm q),\nonumber\\
&\left\{c_2(\bm p);c_2^\dagger(\bm q)\right\}=2p_-\delta(\bm p-\bm q).
\end{align}
And, hence:
\begin{equation}\label{eq:7.1.64}
\left\{u(x);\widetilde u(y)\right\}=-iD_0(x-y).
\end{equation}
The field $\widetilde u(x)$ is called antighost field, and is given by:
\begin{equation}\label{eq:7.1.64.1}
\widetilde u(x)=(2\pi)^{-3/2}\int d\mu(\bm p)\left(-c_1(\bm p)e^{-ipx}+c_2^\dagger(\bm p)e^{ipx}\right).
\end{equation}
Being that way, Eq. \eqref{eq:7.1.62} adopts the final form:
\begin{align}
\left\{Q;Q^\dagger\right\}&=\int d\mu(\bm p)p_-^2\Big(a_-^\dagger(\bm p)a_-(\bm p)+a_+^\dagger(\bm p)a_+(\bm p)\nonumber\\
&+c_1^\dagger(\bm p)c_1(\bm p)+c_2^\dagger(\bm p)c_2(\bm p)\Big).\label{eq:7.1.65}
\end{align}
And now we can affirm:
\begin{equation}\label{eq:7.1.66}
\mathcal F_{\text{phys}}=\text{Ker}\left(\left\{Q;Q^\dagger\right\}\right).
\end{equation}
Also, as a consequence of Eq. \eqref{eq:7.1.60}, and by direct calculation, it is immediate to see that the gauge charge operator is nilpotent:
\begin{equation}\label{eq:7.1.67}
Q^2=\frac{1}{2}\left\{Q;Q\right\}=0.
\end{equation}

The fermion statistics of the ghost particles implies, in Eq. \eqref{eq:7.1.46}, that the parameter $\lambda$ is a number following a Grassmannian multiplication rule, so the series ends in the term linear in $\lambda$: The quantum gauge transformation of the operator $F(x)$, containing gauge, ghosts and anti-ghosts field operators, is \textit{exactly}:
\begin{equation}\label{eq:7.1.68}
F'(x)=e^{-i\lambda Q}F(x)e^{i\lambda Q}=F(x)-i\lambda QF(x)+iF(x)\lambda Q.
\end{equation}
Since $\lambda$ is a Grassmann's number, we cannot simply commute it with $F(x)$ in the last term of Eq. \eqref{eq:7.1.68}; in putting it at the left of $F(x)$ we must write a factor $(-1)^{n_F}$, with $n_F$ the so-called ghost number of the operator $F(x)$, equal to the difference between the number of ghosts and anti-ghost fields contained in it. Therefore, we define the gauge variation of the operator $F(x)$ as:
\begin{equation}\label{eq:7.1.69}
d_QF(x):=\left[Q;F(x)\right]_{(-1)^{1+n_F}}.
\end{equation}
From Eq. \eqref{eq:7.1.68} it follows that the exact gauge transformation is given by:
\begin{equation}\label{eq:7.1.70}
F'(x)=F(x)-i\lambda d_QF(x).
\end{equation}
And from this equation we see that, necessarily, $d_Q$ changes the statistics of the operator $F(x)$, in such a way that $\lambda d_QF(x)$ maintains the one of $F(x)$.

Let us enumerate some properties of the gauge variation $d_Q$. Firstly, it satisfies the following rule for the product of operators:
\begin{equation}\label{eq:7.1.71}
d_Q(F(x)G(y))=(d_QF(x))G(y)+(-1)^{n_F}F(x)(d_QG(y)).
\end{equation}
Secondly, it is nilpotent:
\begin{equation}\label{eq:7.1.73}
d_Q^2=0.
\end{equation}
Thirdly, from Eq. \eqref{eq:7.1.45} they follow the gauge variations of the field operators:
\begin{equation}\label{eq:7.1.74}
d_QA\indices{^a_\pm}=i\partial^a u_\pm,
\end{equation}
\begin{equation}\label{eq:7.1.75}
d_Qu=0,
\end{equation}
\begin{equation}\label{eq:7.1.76}
d_Q\widetilde u_\pm=-i\partial_a A\indices{^a_\pm}.
\end{equation}
And fourthly, it satisfies the following result, proved in Ref. \cite{ScharfGFT}:

\textbf{Lemma:} \textit{Let $G$ be a Wick's monomial containing gauge fields $A$, ghosts $u$ and anti-ghosts $\widetilde u$. The gauge variation of $G$ commutes with its normal ordering:}
\begin{equation}\label{eq:7.1.77}
d_Q\normord{G}=\normord{d_QG}.
\end{equation}

\subsection{Quantum gauge invariance of the scattering operator}

The concept of quantum gauge invariance was already introduced in the commentaries after Eq. \eqref{eq:7.1.43}: It is the independence of the physical quantities from the mathematical operation of extending Fock's space. This kind of symmetries frequently occurs in physics; examples of it are the principle of general covariance, the one of classical gauge symmetry, \textit{et cetera}. In the $S$-matrix program the fundamental physical quantity is the transition amplitude between two physical states. Then \cite{ScharfGFT}:

\textit{Definition.---} Let $P$ be the projection operator onto the physical subspace $\mathcal F_{\text{phys}}$. Two scattering operators $S(g)$ and $S'(g)$ are called physically equivalent if for every states $\Phi,\Psi\in\mathcal F$:
\begin{equation}\label{eq:7.3.1}
\lim_{g\to 1}\left(\Phi;PS(g)P\Psi\right)=\lim_{g\to 1}\left(\Phi;PS'(g)P\Psi\right),
\end{equation}
if the adiabatic limit exists.

The condition of existence of the adiabatic limit is essential, because only in that case the bilinear form $\left(\Phi;PS(g)P\Psi\right)$ has physical meaning; such existence will be assumed in the present discussion. Inserting in this definition the formal series of Eq. \eqref{eq:3.2.3} we obtain a perturbative version of it:

\textit{Definition.---} Two scattering operators $S(g)$ and $S'(g)$ are (perturbatively) physically equivalent at the $n$-th order if their corresponding $n$-point distributions satisfy --the limit is weak in the sense of Eq. \eqref{eq:7.3.1}--:
\begin{equation}\label{eq:7.3.4}
\text{w-}\lim_{g\to 1}\int dX\left(PT_n(X)P-PT'_n(X)P\right)g(X)=0,
\end{equation}
if the adiabatic limit exists for the order $n$ of the perturbation series.

Eq. \eqref{eq:7.3.4} has the general solution $PT_n(X)P=PT'_n(X)P+\partial(\ldots)$, with $\partial(\ldots)$ meaning the divergence of the quantity $(\ldots)$. In effect, the integration by parts implies that in such case the left-hand-side of Eq. \eqref{eq:7.3.4} is equal to:
\begin{equation}\label{eq:7.3.5}
\text{w-}\lim_{g\to 1}\int dX(\ldots)\partial g(X),
\end{equation}
which is null in the adiabatic limit in which $g$ turns to a constant. On the other hand, in the expression $PT_n(X)P$, $T_n(X)$ is defined modulo a gauge variation, because Eq. \eqref{eq:7.1.66} implies that: $PQ=0=QP$, and, as a consequence: $Pd_QXP=PQX-XQP=0$. As a result, we have that the distributions $T_n$ and $T'_n$ are physically equivalent if they differ by terms which are divergences or gauge variations. This we denote by the symbol ``$\sim$'':
\begin{equation}\label{eq:7.3.6}
T'_n\sim T_n\ :\Leftrightarrow\ T'_n= T_n+d_Q(\ldots)+\partial(\ldots).
\end{equation}
Being that way, the matrix elements of the $S(g)$ operator between states of the physical subspace, at any order, are trivially quantum gauge invariant, because $d_Q T_n$ is physically equivalent to the null distribution: $0\sim d_QT_n$. Quantum gauge invariance of $T_1(x)$, therefore, is assured whatever it was. But this is not sufficient: $T_1(x)$ must be constructed in such a way that the second order distribution $T_2(x_1;x_2)$ is also physically equivalent to $T'_2(x_1;x_2)$ constructed with $T'_1(x)=T_1(x)-i\lambda d_QT_1(x)$. This is a consistency condition. And, clearly, the same must be true for the higher order $T_n$ distributions. In a word, we must impose that quantum gauge invariance is not destroyed in the inductive procedure of CPT.

\textit{Definition.---} An interacting theory generated by $T_1(x)$ is a quantum gauge theory if it is physically equivalent to the theory generated by $T_1(x)-i\lambda d_QT_1(x)$ at all orders of the perturbation series.

Let us start by studying the second order distribution, $T_2(x_1;x_2)$. We introduce the following notation: $T_1(x)=T_1^0(x)-i\lambda d_QT_1^0(x)$; $T_2^0(x_1;x_2)$ is the two-points distribution coming from $T_1^0(x)$; $T_2(x_1;x_2)$ is the one corresponding to $T_1(x)$, and so on. As we have already known, the construction of the transition distribution starts by determining the subsidiary distributions:
\begin{align}\label{eq:7.3.7}
&A'_2(x_1;x_2)=-T_1(x_1)T_1(x_2),\nonumber\\
&R'_2(x_1;x_2)=-T_1(x_2)T_1(x_1).
\end{align}
Substituting in them the expression of $T_1(x)$ we have that the second order causal distribution, $D_2=R'_2-A'_2$, is:
\begin{align}
&D_2(x_1;x_2)=D_2^0(x_1;x_2)-i\lambda d_Q D_2^0(x_1;x_2)\nonumber\\
&+2i\lambda T_1^0(x_1)d_QT_1^0(x_2)-2i\lambda T_1^0(x_2)d_QT_1(x_1).\label{eq:7.3.8}
\end{align}
From it we extract the retarded part, $R_2(x_1;x_2)$, by means of the splitting procedure:
\begin{align}
&R_2(x_1;x_2)=R_2^0(x_1;x_2)-i\lambda d_QR_2^0(x_1;x_2)\nonumber\\
&+2i\lambda\text{ret}\left\{T_1^0(x_1)d_QT_1^0(x_2)-T_1^0(x_2)d_QT_1^0(x_1)\right\},\label{eq:7.3.9}
\end{align}
with $\text{ret}\left\{\ldots\right\}$ meaning the retarded part of $\left\{\ldots\right\}$. The two-points distribution $T_2=R_2-R'_2$ is then:
\begin{align}
&T_2(x_1;x_2)=T_2^0(x_1;x_2)-i\lambda d_QT_2^0(x_1;x_2)\nonumber\\
&+2i\lambda\Big(\text{ret}\left\{T_1^0(x_1)d_QT_1^0(x_2)-T_1^0(x_2)d_QT_1^0(x_1)\right\}\nonumber\\
&-T_1^0(x_2)d_QT_1^0(x_1)\Big).\label{eq:7.3.10}
\end{align}
From here it follows that $T_2$ will be physically equivalent to $T_2^0$ if and only if the second line in Eq. \eqref{eq:7.3.10} is a divergence or a gauge variation. The first possibility is realized if $T_1^0(x)$ is a gauge variation: $T_1^0(x)=d_Q(\ldots)$; this case has no physical interest. The second possibility holds when $T_1^0(x)$ is a divergence:
\begin{equation}\label{eq:7.3.11}
d_QT_1^0(x)=i\partial_a T^a_{1/1}(x).
\end{equation}
And, in that way, the theory maintains quantum gauge invariance at second order:
\begin{equation}\label{eq:7.3.12}
T_2\sim T_2^0.
\end{equation}
The $T_{1/1}^a(x)$ distribution is called Q-vertex. Hence, Eq. \eqref{eq:7.3.11} is a necessary condition in every quantum gauge theory. It implies, also, that $T_1(x)$ is not only gauge invariant as a bilinear form in the physical subspace, but in the whole Fock's space.

In order to see if Eq. \eqref{eq:7.3.11} is also a sufficient condition in order to the theory to be quantum gauge invariant we must study the following orders in the perturbation series, which will be easier after establishing the consequences of Eq. \eqref{eq:7.3.11}. As we know, the $n$-point distributions are chronological products of $T_1$ distributions --ordered according to the null-plane time $x^+$--:
\begin{equation}\label{eq:7.3.13}
T_n(x_1;\ldots;x_n)=\mathfrak T_+\left\{T_1(x_1)\ldots T_1(x_n)\right\}.
\end{equation}
Since every $T_1$ has boson character, when we apply the gauge variation operator to $T_n$ we find:
\begin{equation}\label{eq:7.3.14}
d_QT_n=\sum\limits_{l=1}^n\mathfrak T_+\left\{T_1(x_1)\ldots d_QT_1(x_l)\ldots T_1(x_n)\right\}.
\end{equation}
And with Eq. \eqref{eq:7.3.11}:
\begin{equation}\label{eq:7.3.15}
d_QT_n=i\sum\limits_{l=1}^n\partial_a^{x_l}T^a_{n/l}(x_1;\ldots;x_n);
\end{equation}
\begin{equation*}
T^a_{n/l}:=\mathfrak T_+\left\{T_1(x_1)\ldots T_{1/1}^a(x_l)\ldots T_1(x_n)\right\}.
\end{equation*}

Now, $T_1(x)$ is in general a Wick's polynomial:
\begin{equation}\label{eq:7.3.16}
T_1(x)=T_1^0(x)+T_1^1(x)+\ldots+T_1^p(x),
\end{equation}
with every $T_1^j(x)$ a Wick's monomial. Considering Eq. \eqref{eq:7.3.13}, the $T_n(x_1;\ldots;x_n)$ distribution has the general form:
\begin{equation}\label{eq:7.3.17}
T_n(x_1;\ldots;x_n)=\sum\limits_{i_1,\ldots,i_n\in\left\{0;\ldots;p\right\}}T_n^{i_1\ldots i_n}(x_1;\ldots;x_n); 
\end{equation}
with the superindex $i_j$ indicating the one-point distribution $T_1^{i_j}$ corresponding to the point $x_j$, this is to say:
\begin{align}\label{eq:7.3.18}
&T_n^{i_1\ldots i_n}(x_1;\ldots ;x_n):=\nonumber\\
&\quad\mathfrak T_+\left\{T_1^{i_1}(x_1)\ldots T_1^{i_j}(x_j)\ldots T_1^{i_n}(x_n)\right\}.
\end{align}
With this we can already establish the following results, which are general and contain, as a particular case, the sufficiency of Eq. \eqref{eq:7.3.11} as a condition for quantum gauge invariance. We only enunciate them without proof, because they are a generalization of our results at second order and follow with simple modifications from the proof given by D{\"u}tsch in instant dynamics in Ref. \cite{Duetsch1} --see also Ref. \cite{ScharfGFT}--.

\textbf{Theorem:} \textit{Let $T_1(x)$ be a distribution of the form:}
\begin{equation}\label{eq:7.3.19}
T_1(x)=T_1^0(x)+T_1^2(x);\  T_1^2(x)=\partial_a T_1^{4a}(x).
\end{equation}
\textit{If the gauge variation of $T_1^0(x)$ is a divergence,}
\begin{equation}\label{eq:7.3.20}
d_QT_1^0(x)=i\partial_a T_1^{1a}(x),
\end{equation}
\textit{then, in Eq. \eqref{eq:7.3.18} notation:}
\begin{equation}\label{eq:7.3.21}
T_n(x_1;\ldots;x_n)=T_n^{0\ldots 0}(x_1;\ldots;x_n)+\partial(\ldots).
\end{equation}

According to this theorem, any term in $T_1(x)$ which is a divergence can be disregarded without altering the physics at any order in the perturbation series. In particular, this is valid for the divergence of the Q-vertex, which means that Eq. \eqref{eq:7.3.11} is a sufficient condition for quantum gauge invariance at all orders.

\textbf{Theorem:} \textit{Let $T_1(x)$ be a distribution of the form:}
\begin{equation}\label{eq:7.3.30}
T_1(x)=T_1^0(x)+T_1^3(x)\quad;\quad T_1^3(x)=d_QT_1^5(x).
\end{equation}
\textit{If for all order $n$ the gauge variation $T_n^{0\ldots 0}$ is a divergence,}
\begin{equation}\label{eq:7.3.31}
d_QT_n^{0\ldots0}=\partial(\ldots),
\end{equation}
\textit{then $T_n$ and $T_n^{0\ldots0}$ are physically equivalents:}
\begin{equation}\label{eq:7.3.32}
T_n(x_1;\ldots;x_n)=T_n^{0\ldots0}(x_1;\ldots;x_n)+\partial(\ldots)+d_Q(\ldots).
\end{equation}

Note that Eq. \eqref{eq:7.3.31} is always respected once $T_1(x)$ is appropriately chosen [see Eq. \eqref{eq:7.3.15}], \textit{when all the coordinates correspond to different times}. On the other hand, Eq. \eqref{eq:7.3.15} could be invalid only because of instantaneous terms --this is to say, when there are some points corresponding to the same time--, but they can always be cancelled by a convenient choice of the normalization terms. Therefore, all the terms in $T_1(x)$ which are gauge variations can be neglected without any physical implication.

\subsection{Unitarity of the scattering operator in $\mathcal{F}_{\text{phys}}$}

There is, still, one point that must be clarified, which is if the extension of Fock's space and quantum gauge invariance is compatible with the axiom of unitarity of the scattering operator. Remember that such unitarity, being a physical property, must be imposed on $\mathcal F_{\text{phys}}$ only. As this issue has already appeared in our study of Poincar{\'e}'s transformations, let us start by considering Eq. \eqref{eq:7.1.23}, from which it follows that there is a possibility for $\bm{\mathsf U}(a;\Lambda)$ to be pseudo-unitary on the whole Fock's space according to some adjoint ``$K$'', in the sense:
\begin{equation}\label{eq:7.1.32}
\bm{\mathsf U}(a;\Lambda)^K=\bm{\mathsf U}(a;\Lambda)^{-1},
\end{equation}
if the extended quantized field operator $A^a(x)$ is Hermitian according to it: $A^a(x)^K=A^a(x)$. This is:
\begin{align}\label{eq:7.1.33}
&A^a(x)=(2\pi)^{-3/2}\nonumber\\
&\times\sum_\lambda\int d\mu(\bm p)\varepsilon_\lambda(\bm p)^a\left(a_\lambda(\bm p)e^{-ipx}+a_\lambda^K(\bm p)e^{ipx}\right).
\end{align}
Comparison with Eq. \eqref{eq:7.1.9} leads to the equalities:
\begin{equation}\label{eq:7.1.34}
a_\lambda^K(\bm p)=-\sum_\tau g_{\lambda\tau}a_\tau^\dagger(\bm p),
\end{equation}
this is to say: $\quad a_{1,2}^K(\bm p)=a_{1,2}^\dagger(\bm p)$ and $a_\pm^K(\bm p)=-a_\mp^\dagger(\bm p)$. In instant dynamics with the physical space containing photon wavefunctions in the radiation gauge, one can find explicitly that a similar adjoint exists by using the definition $B^K=(-1)^{\mathsf N_0}B^\dagger(-1)^{\mathsf N_0}$, with $\mathsf N_0$ the number of scalar particles operator \cite{ScharfFQED,ScharfGFT}. In light-front dynamics, instead, while not finding its explicit expression, we will prove that such an adjoint ``$K$'' does exist, by considering the following theorem \cite{Prugovecki}:

\textbf{Theorem:} \textit{Let $(\bullet;\bullet)_K$ be a bilinear form in Hilbert's space $\mathcal{H}$. If it is bounded,} i.e.\textit{:}
\begin{equation}\label{eq:7.1.34-1}
\exists C\in\mathbb{R}_+:\ \forall f,g\in\mathcal{H}:\  \left|(f;g)_K\right|\leq C\|f\|\|g\|\ ,
\end{equation}
\textit{then there is a bounded linear operator $\eta:\mathcal{H}\to\mathcal{H}$, with $\|\eta\|\leq C$ and $\text{Dom}(\eta)=\mathcal{H}$, such that:}
\begin{equation}\label{eq:7.1.34-2}
(f;g)_K=(f;\eta g).
\end{equation}
\textit{Moreover, $(\bullet;\bullet)_K$ is Hermitian}\footnote{This is to say, for every two vectors $f,g\in\mathcal{H}$: $(f;g)_K=(g;f)^*_K$.} \textit{if and only if the operator $\eta$ is self-adjoint, $\eta^\dagger=\eta$.}

We start by defining the following operators, for test functions $f\in\mathscr S(\mathbb{R}^3)$:
\begin{align}\label{eq:7.1.34-3}
&a^a_\lambda(f):=\int d\mu(\bm p)\hat f(\bm p)^*\varepsilon_\lambda(\bm p)^aa_\lambda(\bm p),\nonumber\\
&a^a_\lambda(f)^\dagger:=\int d\mu(\bm p)\hat f(\bm p)\varepsilon_\lambda(\bm p)^aa_\lambda^\dagger(\bm p).
\end{align}
They generate the states:
\begin{equation}\label{eq:7.1.34-4}
\Phi^{a_1\cdots a_n}_{\lambda_1\cdots\lambda_n}(f_1;\cdots;f_n):=a^{a_1}_{\lambda_1}(f_1)^\dagger\cdots a^{a_n}_{\lambda_n}(f_n)^\dagger\Omega.
\end{equation}
Similarly, having Eq. \eqref{eq:7.1.34} as inspiration, we define the operators:
\begin{equation}\label{eq:7.1.34-5}
a^a_\lambda(f)^K:=-\int d\mu(\bm p)\hat f(\bm p)\varepsilon_\lambda(\bm p)^a\sum_\tau g_{\tau\lambda}a_\tau^\dagger(\bm p),
\end{equation}
and the states generated by them are denoted:
\begin{equation}\label{eq:7.1.34-6}
\widetilde\Phi^{a_1\cdots a_n}_{\lambda_1\cdots\lambda_n}(f_1;\cdots;f_n):=a^{a_1}_{\lambda_1}(f_1)^K\cdots a^{a_n}_{\lambda_n}(f_n)^K\Omega.
\end{equation}
We define now the bilinear form:
\begin{align}\label{eq:7.1.34-7}
&\left(\Phi_{\lambda_1\cdots\lambda_n}(f_1;\cdots;f_n);\Phi_{\sigma_1\cdots\sigma_m}(g_1;\cdots;g_m)\right)_K:=\nonumber\\
&\quad\left(\Phi_{\lambda_1\cdots\lambda_n}(f_1;\cdots;f_n);\widetilde\Phi_{\sigma_1\cdots\sigma_m}(g_1;\cdots;g_m)\right).
\end{align}
This definition, given only for the states of Eq. \eqref{eq:7.1.34-4}, is suficient because they generate Fock's space and because it is defined with regard to the inner product; its evaluation for more general states is defined by linear continuation.

Attending at first at the one-particle states, the direct calculus shows that:
\begin{align}\label{eq:7.1.34-8}
&\left(\Phi_\lambda(f);\Phi_\sigma(g)\right)_K=\nonumber\\
&-g_{\lambda\sigma}\int d\mu(\bm p)\hat f(\bm p)^*\hat g(\bm p)\left(\sum_ a\varepsilon_\lambda(\bm p)^a\varepsilon_\sigma(\bm p)^a\right).
\end{align}
By comparison, the inner product between the same states is:
\begin{align}\label{eq:7.1.34-9}
&\left(\Phi_\lambda(f);\Phi_\sigma(g)\right)=\nonumber\\
&\delta_{\lambda\sigma}\int d\mu(\bm p)\hat f(\bm p)^*\hat g(\bm p)\left(\sum_ a\varepsilon_\lambda(\bm p)^a\varepsilon_\sigma(\bm p)^a\right).
\end{align}
So we see that for the physical polarizations $\lambda=1,2$, the bilinear form $(\bullet;\bullet)_K$ reduces to the inner product, which satisfies Eq. \eqref{eq:7.1.34-1} with $C=1$ because of Cauchy-Schwarz's inequality. The other case in which the bilinear product is non-null (modulo symmetry operations) is:
\begin{align}
&\left(\Phi_+(f);\Phi_-(g)\right)_K=-\int d\mu(\bm p)\hat f(\bm p)^*\hat g(\bm p)\nonumber\\
&\qquad\quad\times\left(\sum_ a\varepsilon_+(\bm p)^a\varepsilon_-(\bm p)^a\right)\nonumber\\
&\qquad=-\int d\mu(\bm p)\frac{p_\perp^2}{2p_-^2}\hat f(\bm p)^*\hat g(\bm p),\label{eq:7.1.34-10}
\end{align}
in which we have used the explicit form of the polarization vectors. On the other hand, from Eq. \eqref{eq:7.1.34-10} we obtain:
\begin{align}\label{eq:7.1.34-11}
&\left\|\Phi_+(f)\right\|^2=\int d\mu(\bm p)\left(1+\frac{p_\perp^2}{2p_-^2}\right)^2\left|\hat f(\bm p)\right|^2,\nonumber\\
&\left\|\Phi_-(g)\right\|^2=\int d\mu(\bm p)\left|\hat g(\bm p)\right|^2.
\end{align}
Eqs. \eqref{eq:7.1.34-10} and \eqref{eq:7.1.34-11}, together with Cauchy-Schwarz's inequality in $L^2(\mathcal{M},\mu)$, imply that:
\begin{align}
\left|\left(\Phi_+(f);\Phi_-(g)\right)_K\right|&=\left|\int d\mu(\bm p)\left(\frac{p_\perp^2}{2p_-^2}\hat f(\bm p)\right)^*\hat g(\bm p)\right|\nonumber\\
&=\left|\left(\frac{p_\perp^2}{2p_-^2}\hat f(\bm p);\hat g(\bm p)\right)_{L^2}\right|\nonumber\\
&\leq \left\|\frac{p_\perp^2}{2p_-^2}\hat f\right\|\left\|\hat g\right\|\nonumber\\
&\leq\left\|\Phi_+(f)\right\|\left\|\Phi_-(g)\right\|.\label{eq:7.1.34-12}
\end{align}
Therefore, also in this case Eq. \eqref{eq:7.1.34-1} is satisfied with $C=1$. Now we can generalize to the case in which the states are linear combinations of different polarizations, as for the same polarization: $\Phi^a_{\lambda}(f_1)+\Phi^a_\lambda(f_2)=\Phi^a_\lambda(f_1+f_2)$. Hence, since each polarization has only one corresponding polarization such that the bilinear form is non-null, as indicated by Eq. \eqref{eq:7.1.34-8}, we will have, for example:
\begin{align*}
&\left|\left(\Phi_{\lambda_1}+\Phi_{\lambda_2};\Phi_{\sigma_1}+\Phi_{\sigma_2}\right)_K\right|\nonumber\\
&\quad=\left|\left(\Phi_{\lambda_1};\Phi_{\sigma_1}\right)_K+\left(\Phi_{\lambda_2};\Phi_{\sigma_2}\right)_K\right|\\
&\quad\leq\left|\left(\Phi_{\lambda_1};\Phi_{\sigma_1}\right)_K\right|+\left|\left(\Phi_{\lambda_2};\Phi_{\sigma_2}\right)_K\right|\\
&\quad\leq\left\|\Phi_{\lambda_1}\right\|\left\|\Phi_{\sigma_1}\right\|+\left\|\Phi_{\lambda_2}\right\|\left\|\Phi_{\sigma_2}\right\|\\
&\quad\leq\left(\left\|\Phi_{\lambda_1}\right\|^2+\left\|\Phi_{\lambda_2}\right\|^2\right)^{1/2}\left(\left\|\Phi_{\sigma_1}\right\|^2+\left\|\Phi_{\sigma_2}\right\|^2\right)^{1/2}.
\end{align*}
The last line contains the norm of the states $\Phi^a_{\lambda_1}+\Phi^a_{\lambda_2}$ and $\Phi^b_{\sigma_1}+\Phi^b_{\sigma_2}$, because the polarization vectors --and then the corresponding states-- are mutually orthogonal --see Eq. \eqref{eq:7.1.34-9}--. The same applies when other polarizations are present, so even in the most general linear combination:
\begin{equation}\label{eq:7.1.34-13}
\left|\left(\Phi;\Psi\right)_K\right|\leq\|\Phi\|\|\Psi\|.
\end{equation}
As the constant is $C=1$, its value will not be modified when tensor products are considered, so Eq. \eqref{eq:7.1.34-13} is valid for any $\Phi,\Psi\in\mathcal{F}$. The above theorem then implies that there is an operator $\eta$, defined over the entire Fock's space, such that:
\begin{equation}\label{eq:7.1.34-14}
(\Phi;\Psi)_K=(\Phi;\eta\Psi),\ \|\eta\|\leq 1.
\end{equation}
Even more, from Eq. \eqref{eq:7.1.34-8} it is clear that the bilinear form $(\bullet;\bullet)_K$ is Hermitian,
\begin{equation}\label{eq:7.1.34-15}
\left(\Phi_\lambda(f);\Phi_\sigma(g)\right)^*_K=\left(\Phi_\sigma(g);\Phi_\lambda(f)\right)_K,
\end{equation}
from which it follows that the operator $\eta$ is self-adjoint:
\begin{equation}\label{eq:7.1.34-16}
\eta^\dagger=\eta.
\end{equation}
Utilizing Eqs. \eqref{eq:7.1.34-4} and \eqref{eq:7.1.34-6} it is possible to show that:
\begin{equation}\label{eq:7.1.34-17}
\left(\widetilde\Phi_\lambda(f);\widetilde\Phi_\sigma(g)\right)=\left(\Phi_\lambda(f);\Phi_\sigma(g)\right),
\end{equation}
while Eqs. \eqref{eq:7.1.34-14}-\eqref{eq:7.1.34-16} imply:
\begin{align}
&\left(\widetilde\Phi_\lambda(f);\widetilde\Phi_\sigma(g)\right)=\left(\widetilde\Phi_\lambda(f);\Phi_\sigma(g)\right)_K=\left(\widetilde\Phi_\lambda(f);\eta\Phi_\sigma(g)\right)\nonumber\\
&\quad=\left(\eta\Phi_\sigma(g);\widetilde\Phi_\lambda(f)\right)^*=\left(\eta\Phi_\sigma(g);\Phi_\lambda(f)\right)^*_K\nonumber\\
&\quad=\left(\eta\Phi_\sigma(g);\eta\Phi_\lambda(f)\right)^*=\left(\Phi_\lambda(f);\eta^2\Phi_\sigma(g)\right).\label{eq:7.1.34-18}
\end{align}
The comparison of Eqs. \eqref{eq:7.1.34-17} and \eqref{eq:7.1.34-18} leads to establish that:
\begin{equation}\label{eq:7.1.34-19}
\eta^2=1.
\end{equation}

Other important properties are the following. Since the bilinear form $(\bullet;\bullet)_K$ reduces to the inner product between physical states --transverse polarizations--, we will have that:
\begin{equation}\label{eq:7.1.34-20}
\forall\Phi\in\mathcal{F}_{\text{phys}}:\ \eta\Phi=\Phi.
\end{equation}
This is valid, particularly, for the vacuum state: $\eta\Omega=\Omega$. Also, from Eqs. \eqref{eq:7.1.34-7} and \eqref{eq:7.1.34-14}:
\begin{equation}\label{eq:7.1.34-21}
\left(\Phi;\eta a^\dagger_{\lambda_1}(f_1)\cdots a^\dagger_{\lambda_n}(f_n)\Omega\right)=\left(\Phi;a^K_{\lambda_1}(f_1)\cdots a^K_{\lambda_n}(f_n)\Omega\right).
\end{equation}
Introducing the identity $1=\eta^2$ between any two emission operators in the left-hand-side of this equation, we arrive at the rigorous definition of the adjoint ``$K$'': Let $B$ be an operator:
\begin{equation}\label{eq:7.1.34-22}
B^K:=\eta B^\dagger\eta.
\end{equation}
This definition and the properties of the operator $\eta$ found above allow one to show that the adjoint ``$K$'' is an involution:
\begin{equation*}
(A+B)^K=A^K+B^K,\ (AB)^K=B^KA^K,
\end{equation*}
\begin{equation}
(B^K)^K=B,\ (cB)^K=c^*B^K\ (c\in\mathbb{C}).\label{eq:7.1.34-23}
\end{equation}
In this way, we have shown that there is an adjoint ``$K$'' such that the extended radiation field is pseudo-Hermitian, and accordingly, such that the operator $\bm{\mathsf U}(a;\Lambda)$ is pseudounitary in the entire Fock's space.

Finally, in order that the pseudounitarity according to ``$K$'' to be maintained even after a quantum gauge transformation as in Eq. \eqref{eq:7.1.46}, one needs to impose also the pseudo-Hermiticity of $u(x)$:
\begin{equation}\label{eq:7.4.1}
c_1^K(\bm p)=c_2^\dagger(\bm p)\ ,\  c_2^K(\bm p)=c_1^\dagger(\bm p).
\end{equation}
With this it follows that the antighost field is anti-pseudo-Hermitian under ``$K$'': $\widetilde u(x)^K=-\widetilde u(x)$. Also, the ghost fields can be written now as:
\begin{equation}\label{eq:7.4.2}
u(x)=(2\pi)^{-3/2}\int d\mu(\bm p)\left(c_2(\bm p)e^{-ipx}+c_2^K(\bm p)e^{ipx}\right); 
\end{equation}
\begin{equation}\label{eq:7.4.3}
\widetilde u(x)=(2\pi)^{-3/2}\int d\mu(\bm p)\left(-c_1(\bm p)e^{-ipx}+c_1^K(\bm p)e^{ipx}\right).
\end{equation}

The following result is immediate --its proof is identical to the proof of unitarity in the physical Fock's space (see Ref. \cite{APS22})--.

\textbf{Theorem:} \textit{Let $T_1(x)$ be the one-point distribution of a quantum gauge theory. If $T_1$ satisfies the perturbative pseudo-unitarity condition,}
\begin{equation}\label{eq:7.4.4}
\widetilde T_1(x)=T_1(x)^K,
\end{equation} 
\textit{then the $n$-point distributions can be constructed, by the inductive procedure of CPT, so as to satisfy the same pseudo-unitarity condition.}

Since the adjoints ``$K$'' and ``$\dagger$'' coincide in $\mathcal{F}_{\text{phys}}$, the above theorem implies that the unitarity in the physical subspace is not destroyed in the extension of Fock's space process.

To finish this section, we want to say a word about the bilinear form $(\bullet;\bullet)_K$. It corresponds to the ``undefinite metric'' used in Gupta-Bleuler's quantization procedure. If one adopts it as the ``inner product'' (and therefore uses pseudo-Hilbert or Krein spaces), the non-transverse states have zero norm, as can be seen from Eq. \eqref{eq:7.1.34-8}. This is similar to what happens in instant dynamics, but in that case the metrics is $\text{diag}(+1;-1;-1-1)$, hence one non-transverse mode has negative norm while the other has a positive one. We think that it is very important to show that pseudo-Hilbert spaces are unnecessary in the sense that there is no reason to say that the bilinear form $(\bullet;\bullet)_K$ is the inner product: The extended Fock's space is a perfectly defined Hilbert space and the bilinear form $(\bullet;\bullet)_K$, which is not an inner product, simply defines a different involution, which coincides with the ``$\dagger$'' adjoint in the physical subspace. In this sense, both physics and mathematics are put on a solid and safe ground.

\section{Ward-Takahashi's identities for QED}\label{sec:WTI}

Now we turn to the application of the general results just founded to null-plane QED. The procedure to obtain Ward-Takahashi's identities is as in the instant dynamics formulation \citep{ScharfGFT}. Recall \cite{APQED1} that the one-point distribution for QED only contains the interaction between the gauge field and the matter fields:
\begin{equation}\label{eq:8.0.14}
T_1(x)=iej^a(x)A_a(x),
\end{equation}
with $j^a(x)$ the matter fields current. In that case, the gauge variation of $T_1$ is equal to
\begin{align}\label{eq:8.0.14-1}
d_QT_1(x)&=-ej^a(x)\partial_au(x)\nonumber\\
&=\partial_a\left(-ej^a(x)u(x)\right)+e\left(\partial_a j^a(x)\right)u(x),
\end{align}
which reduces to a divergence if the current $j^a$ is conserved; in such case, the second term in the above equation is null. Hence: The field-current coupling in CPT holds with the conserved current of the free matter fields. For a fermion field, it is:
\begin{equation}\label{eq:8.0.15}
j^a(x)=\normord{\overline\psi(x)\gamma^a\psi(x)}.
\end{equation}
The Q-vertex for QED is thus:
\begin{equation}\label{eq:8.0.16}
T_{1/1}^a(x)=iej^a(x)u(x)=ie\normord{\overline\psi(x)\gamma^a\psi(x)}u(x).
\end{equation}

Now we want to explore the consequences of gauge invariance in the higher order terms $T_n$. Since there is no selfinteraction term of the radiation field in $T_1$, at each point only a gauge field can appear, and the $T_n$ distribution will have the general form:
\begin{equation}\label{eq:8.1.1}
T_n(x_1;\cdots;x_n)=\normord{T_l^a(x_1;\cdots;x_n)A_a(x_l)}+\cdots,
\end{equation}
in which neither $T_l^a$ nor the ellipsis include another gauge operator at the point $x_l$, $A_a(x_l)$. Being that way, the gauge variation of $T_n$ is:
\begin{equation}\label{eq:8.1.2}
d_QT_n(x_1;\cdots;x_n)=i\normord{T_l^a(x_1;\cdots;x_n)\partial_a^{x_l}u(x_l)}+\cdots,
\end{equation}
and this time the ellipsis does not include any ghost field at the point $x_l$, $u(x_l)$. According to Eq. \eqref{eq:7.3.15}, which is the general condition for gauge invariance at order $n$, the gauge variation in Eq. \eqref{eq:8.1.2} must reduce to a divergence. Leibniz's rule implies that this is possible --note that this procedure is identical to that in Eq. \eqref{eq:8.0.14-1}--, and
\begin{equation}\label{eq:8.1.3}
d_QT_n(x_1;\cdots;x_n)=i\partial_a^{x_l}\left(\normord{T^a_l(x_1;\cdots;x_n)u(x_l)}\right)+\cdots,
\end{equation}
whenever the Ward-Takahashi's identities hold:
\begin{equation}\label{eq:8.1.4}
\partial_a^{x_l}\normord{T^a_l(x_1;\cdots;x_n)}=0.
\end{equation}
These equations are valid at every point $x_l$ at which there is a radiation field. If there is no radiation field at a given point, the gauge variation will be automatically null, and the corresponding term, gauge invariant.

Particularly, application of Ward-Takahashi's identity to vacuum polarization leads to the condition of transversality to the momentum, which in real space is expressed as
\begin{equation*}
\partial_a^{x_1}\Pi^{ab}(x_1;x_2)=0=\partial_b^{x_2}\Pi^{ab}(x_1;x_2),
\end{equation*}
and which has been implicitly used in our previous paper, Ref. \cite{APQED1}, in order to normalize the vacuum polarization scalar only, leaving unchanged the tensor structure.

\section{Electron's self-energy}\label{sec:self}

In this section we will use the obtained results to calculate electron's self-energy, using only the covariant part of the radiation field commutation distribution. As was shown, such restriction has no influence in the matrix elements of the scattering operator between physical states. Electron's self-energy comes from the causal distribution (see Ref. \cite{APQED1}):
\begin{align}\label{eq:8.6.1}
D_2^{(SE)}(x_1;x_2)=&\normord{\overline\psi(x_1)d(y)\psi(x_2)}\nonumber\\
&-\normord{\overline\psi(x_2)d(-y)\psi(x_1)},
\end{align}
with:
\begin{align}\label{eq:8.6.2}
&d(y):=-e^2\gamma^a\left[d^+(y)+d^-(y)\right]\gamma_a;\nonumber\\
& d^\pm(y):=S_{\pm}(y)D_{0+}(\pm y).
\end{align}
We will start by calculating Fourier's transform of $d^-$, which is:
\begin{align}
\hat d^-(p)&=(2\pi)^{-2}\int d^4q\widehat S_-(p-q)\widehat D_{0+}(-q)\nonumber\\
&=(2\pi)^{-4}\left[(\slashed p+m)I_1-\gamma^aI_{2a}\right],\label{eq:8.6.3}
\end{align}
with the following definitions of the various integrals:
\begin{equation}\label{eq:8.6.4}
I_1=\int d^4q\Theta(-q_-)\Theta(q_--p_-)\delta(q^2)\delta\left((p-q)^2-m^2\right),
\end{equation}
\begin{equation}\label{eq:8.6.5}
I_{2a}=\int d^4q\Theta(-q_-)\Theta(q_--p_-)\delta(q^2)\delta\left((p-q)^2-m^2\right)q_a.
\end{equation}

The calculus of these integrals is simplified in the reference frame in which $p=(p_+;0_\perp;p_-)$, that exists because $q\in\overline{V^-}$ and $p-q\in V^-$, so that $p=q+(p-q)\in V^-$; this implies, in particular, that $p_+,p_-<0$. In that reference frame:
\begin{equation}\label{eq:8.6.8}
\left.(p-q)^2-m^2\right|_{q^2=0}=2p_+p_--2p_+q_--2p_-q_+-m^2.
\end{equation}
Then, using the properties of Dirac's delta distributions:
\begin{align}\label{eq:8.6.9}
&\Theta(-q_-)\delta(q^2)\delta\left((p-q)^2-m^2\right)=\frac{1}{|2p_-|}\Theta(-q_-)\nonumber\\
&\times\Theta(q_--A)\Theta(2p_+p_--m^2)\delta\left(q_+-\frac{q_\perp^2}{2q_-}\right)\nonumber\\
&\times\delta\left[q_\perp^2-\frac{2p_+}{p_-}(A-q_-)q_-\right],
\end{align}
with:
\begin{equation}\label{eq:8.6.10}
A=\frac{2p_+p_--m^2}{2p_+}.
\end{equation}
Since $p_+<0$, it is trivial that $A>p_-$, so that Heaviside's function $\Theta(q_--p_-)$ appearing in the integrals of Eqs. \eqref{eq:8.6.4} and \eqref{eq:8.6.5} is redundant. Also, in the chosen reference frame it is clear, by symmetry arguments, that:
\begin{equation}\label{eq:8.6.11}
I_{2\alpha}=0.
\end{equation}
Then, the only integrals that we need to calculate are: $I_1$, $I_{2+}$ and $I_{2-}$. The integration in the variables $q_+$ and $q_\perp^2$ is immediate by using the supports of Dirac's delta distributions in Eq. \eqref{eq:8.6.9}; we obtain:
\begin{equation}\label{eq:8.6.13}
I_1=\frac{\pi}{|2p_-|}\Theta(-p_-)\Theta(2p_+p_--m^2)\int\limits_A^0dq_-,
\end{equation}
\begin{equation}\label{eq:8.6.14}
I_{2+}=\frac{\pi}{|2p_-|}\Theta(-p_-)\Theta(2p_+p_--m^2)\frac{p_+}{p_-}\int\limits_A^0dq_-(A-q_-),
\end{equation}
\begin{equation}\label{eq:8.6.15}
I_{2-}=\frac{\pi}{|2p_-|}\Theta(-p_-)\Theta(2p_+p_--m^2)\int\limits_A^0dq_-q_-,
\end{equation}
The integration in the $q_-$ variable is now elementary:
\begin{equation}\label{eq:8.6.19}
I_1=\frac{\pi}{2}\Theta(-p_-)\Theta(2p_+p_--m^2)\left(1-\frac{m^2}{2p_+p_-}\right),
\end{equation}
\begin{equation}\label{eq:8.6.20}
I_{2\pm}=\frac{p_\pm}{2}\left(1-\frac{m^2}{2p_+p_-}\right)I_1.
\end{equation}
Substituting these results into Eq. \eqref{eq:8.6.3} and multiplying by $\gamma^a$ by the left and by $\gamma_a$ by the right we obtain:
\begin{align}
\gamma^a&\hat d^-(p)\gamma_a=(2\pi)^{-3}\Theta(-p_-)\Theta(2p_+p_--m^2)\nonumber\\
&\times\left(1-\frac{m^2}{2p_+p_-}\right)\left\{m-\frac{\slashed p}{4}\left(1+\frac{m^2}{2p_+p_-}\right)\right\}.\label{eq:8.6.24}
\end{align}

In an analogous manner, it can be found:
\begin{align}
\gamma^a&\hat d^+(p)\gamma_a=-(2\pi)^{-3}\Theta(p_-)\Theta(2p_+p_--m^2)\nonumber\\
&\times\left(1-\frac{m^2}{2p_+p_-}\right)\left\{m-\frac{\slashed p}{4}\left(1+\frac{m^2}{2p_+p_-}\right)\right\}.\label{eq:8.6.25}
\end{align}
Replacing Eqs. \eqref{eq:8.6.24} and \eqref{eq:8.6.25} into Eq. \eqref{eq:8.6.2} we finally arrive at the causal distribution:
\begin{align}
\hat d(p)=&e^2(2\pi)^{-3}\sgn(p_-)\Theta(2p_+p_--m^2)\left(1-\frac{m^2}{2p_+p_-}\right)\nonumber\\
&\times\left\{m-\frac{\slashed p}{4}\left(1+\frac{m^2}{2p_+p_-}\right)\right\}.\label{eq:8.6.26}
\end{align}

Because of the polynomial factorization theorem \cite{APQED1}, in order to obtain its retarded part we write it in the following form:
\begin{align}\label{eq:8.6.27}
\hat d(p)=&e^2(2\pi)^{-3}(2p_+p_--m^2)\nonumber\\
&\times\left\{2p_+p_-m-\frac{\slashed p}{4}(2p_+p_-+m^2)\right\}\hat d_1(p),
\end{align}
with:
\begin{equation}\label{eq:8.6.28}
\hat d_1(p)=\sgn(p_-)\Theta(2p_+p_--m^2)\frac{1}{(2p_+p_-)^2}.
\end{equation}
Now we only need to split this distribution, whose singular order at the $x^-$ axis is negative \footnote{The reason for this value of the singular order is the integral
\begin{equation*}
\int\rho(s)\Theta\left(\frac{p}{s}-a\right)\frac{s^2}{p^2}\hat f(p)dp=s\rho(s)\int\limits_a^{+\infty}\hat f(sq)\frac{dq}{q^2},
\end{equation*}
which goes to $\hat f(0)/a$ as $s\to 0$ provided we choose the function $\rho(s)=s^{-1}$ --this defines the value $\omega_-=-1$ of the distribution $\hat d_1$--. In that case, in the limit of $s\to 0$:
\begin{equation*}
\rho(s)\Theta\left(\frac{p}{s}-a\right)\frac{s^2}{p^2}\to \frac{\delta(p)}{a}.
\end{equation*}
This is a departure from power-counting, which assigns a value $-2$ for the degree of $d_1$. This departure is irrelevant in this case because the formula for obtaining the retarded distribution is the same for all negative values of the singular order. Nonetheless, to determine the correct value of the singular order is extremely important as it can lead to different physical results if the singular order is non-negative \cite{AsteSO}.}:
\begin{equation}\label{eq:8.6.30}
\omega_-\left[d_1\right]=-1.
\end{equation}
Accordingly, its splitting is done via Eq. \eqref{eq:3.6.9}. Performing the change $s=-2kp_-+2p_+p_-$:
\begin{align}
\hat r_1(p)&=\frac{i}{2\pi}\int\frac{dk}{k+i0^+}\sgn(p_-)\Theta(-2kp_-+2p_+p_--m^2)\nonumber\\
&\quad\times\frac{1}{(-2kp_-+2p_+p_-)^2}\nonumber\\
&=\frac{i}{2\pi}\Bigg\{\text{V.p.}\int\limits_{m^2}^{+\infty}\frac{ds}{s^2(2p_+p_--s)}\nonumber\\
&\quad-i\pi\sgn(p_-)\Theta(2p_+p_--m^2)\frac{1}{(2p_+p_-)^2}\Bigg\}\nonumber\\
&=\frac{i}{2\pi}\frac{1}{(2p_+p_-)^2}\Bigg\{\log\left(\left|\frac{2p_+p_--m^2}{m^2}\right|\right)+\frac{2p_+p_-}{m^2}\nonumber\\
&\quad-i\pi\sgn(p_-)\Theta(2p_+p_--m^2)\Bigg\}.\label{eq:8.6.31}
\end{align}

Therefore we obtain the retarded distribution:
\begin{align}
&\hat r(p)=\frac{ie^2}{(2\pi)^4}\Bigg\{\left(1-\frac{m^2}{2p_+p_-}\right)\left[m-\frac{\slashed p}{4}\left(1+\frac{m^2}{2p_+p_-}\right)\right]\nonumber\\
&\times\left[\log\left(\left|\frac{2p_+p_--m^2}{m^2}\right|\right)-i\pi\sgn(p_-)\Theta(2p_+p_--m^2)\right]\nonumber\\
&+\frac{m^2\slashed p}{4(2p_+p_-)}-m-\frac{2p_+p_-}{m}(\slashed p-m)\Bigg\}.\label{eq:8.6.37}
\end{align}
In this expression, the last two terms in the second line have the form of normalization terms, so they can be replaced by arbitrary values, having in mind that the singular order at the $x^-$ axis of the complete causal distribution in Eq. \eqref{eq:8.6.26} is $\omega_-\left[d\right]=+1$. Also, the subsidiary retarded distribution is $\hat r'(p)=-e^2\gamma^a\hat d^-(p)\gamma_a$; its value is given in Eq. \eqref{eq:8.6.24}. Therefore, if we define the fermion self-energy $\Sigma$ such that the transition distribution is:
\begin{align}\label{eq:8.6.38}
T_2^{(SE)}&(x_1;x_2)=i\normord{\overline{\psi}(x_1)\Sigma(x_1-x_2)\psi(x_2)}\nonumber\\
&+i\normord{\overline\psi(x_2)\Sigma(x_2-x_1)\psi(x_1)},
\end{align}
then:
\begin{equation}\label{eq:8.6.39}
\widehat\Sigma(p)=-i\hat t_2^{(SE)}(p)=-i\left(\hat r(p)-\hat r'(p)\right),
\end{equation}
so that ($C_0$ and $C_1$ are normalization constants), in Lorentz's covariant form we finally get:
\begin{align}
\widehat\Sigma(p)=&\frac{e^2}{(2\pi)^4}\Bigg\{\left(1-\frac{m^2}{p^2}\right)\left[m-\frac{\slashed p}{4}\left(1+\frac{m^2}{p^2}\right)\right]\nonumber\\
&\left[\log\left(\left|\frac{p^2-m^2}{m^2}\right|\right)-i\pi\Theta(p^2-m^2)\right]\nonumber\\
&+\frac{m^2\slashed p}{4p^2}+C_0+C_1\slashed p\Bigg\}.\label{eq:8.6.40}
\end{align}
This distribution is the same of that in instant dynamics \cite{ScharfFQED}.

In order to fix the normalization constants $C_0$ and $C_1$ we study electron's self-energy insertions into Compton's scattering, following the path shown for Yukawa's model in Ref. \cite{AP2}; we find that the complete fermion propagator is the one which solves the equation:
\begin{equation}\label{eq:8.6.41}
\widehat S_{\text{tot}}=\hat t_2^{(C)}\left(1+(2\pi)^4\widetilde\Sigma\widehat S_{\text{tot}}\right),
\end{equation}
with $\hat t_2^{(C)}$ the normalized fermion Feynman's propagator (without instantaneous term), as it is the one that appear in the transition distribution for Compton's scattering at second order \cite{APQED1}:
\begin{equation}\label{eq:8.6.42}
\hat t_2^{(C)}(p)=(2\pi)^{-2}\frac{1}{\slashed p-m+i0^+}.
\end{equation}
The solution is:
\begin{equation}\label{eq:8.6.43}
\widehat S_{\text{tot}}(p)=(2\pi)^{-2}\frac{1}{\slashed p-\left(m+(2\pi)^2\widetilde\Sigma(p)\right)+i0^+}.
\end{equation}
Then one imposes the normalization conditions: (1) The parameter $m$ is the fermion's physical mass; (2) the parameter $e$ is the physical value of the electric charge. These conditions, respectively, are translated into:
\begin{equation}\label{eq:8.6.44}
\lim_{\slashed p\to m}\widetilde\Sigma(p)=0,\ \lim_{\slashed p\to m}\frac{d\widetilde\Sigma(p)}{d\slashed p}=0.
\end{equation}
The first of them can be directly imposed, since $\widetilde\Sigma$ in Eq. \eqref{eq:8.6.40} does not diverge on the mass shell; it leads to the relation:
\begin{equation}\label{eq:8.6.45}
C_0=-m\left(C_1+\frac{1}{4}\right).
\end{equation}
The second normalization condition, however, cannot be directly imposed, as the derivative of $\widetilde\Sigma$ is singular on the mass shell. Such infrared problem is present in every formalism: non-analyticity is a general property when a massless distribution is part of the convolution (when an internal line corresponds to a massless particle) \cite{Bogo}. There are some partial solutions to this problem in the literature, \textit{e.g.}, to introduce a non-null mass for the photon \cite{Bogo, ItzyksonZuber}, or to use Pauli-Villars' regularization \cite{WeinbergQFT1}; the dependence on the unphysical parameters, however, cannot be posteriorly eliminated. Fortunately, one can avoid such complications by using Ward-Takahashi's identities to show that the normalization of self-energy and that of the vertex function are not independent: they are related in such a way that the normalization of one of them is compensated by the normalization of the other so that the electric charge is not changed. Because of this, the normalization constant $C_1$ does not contain any physics and can be chosen to satisfy a different normalization condition in which infrared divergences are not present. For example, normalizing at $p=0$:
\begin{equation}\label{eq:8.6.46}
\lim_{\slashed p\to 0}\frac{d\widetilde\Sigma(p)}{d\slashed p}=0\ \Rightarrow\  C_1=\frac{1}{8}.
\end{equation}

\section{Conclusions}\label{sec:Conc}

We have shown an explicit construction of a quantum gauge invariant theory in light-front dynamics, which proves the independence of the physical scattering operator of the gauge terms in the massless gauge field commutation distribution. For such a task it was necessary to extend Fock's space to contain also non-physical states, not only coming from the non-physical polarization states of the gauge field, but also from massless fermion ghost fields. The scattering operator matrix elements between physical states are then independent of the gauge terms of the commutation distribution whenever the normalization terms in the splitting of the causal distribution are chosen so as to satisfy Ward-Takahashi's identities. In particular, this imposes a restriction over the one-point distribution $T_1$, similarly as classical gauge invariance dictates the form of the Lagrangian interaction density by Utiyama's minimal coupling prescription.

With the aid of the result just commented, we have shown (see the appendix) the normalizability of the physical scattering matrix of null-plane QED, as well as that no selfinteraction terms for the fields will appear at any order. Additionally, we have calculated electron's self-energy by utilizing the covariant part of the radiation field commutation distribution, showing by direct comparison the equivalence with instant dynamics. Its normalization has been done by studying its insertions into Compton's scattering. In that way, a relation between the two undetermined constants which appear in its expression follows from the imposition of the physical value of the mass of the lepton. On the other hand, the physical value of the coupling constant cannot be imposed on the mass shell because of the infrared divergences, hence one imposes a condition at a different normalization point. To show that such a change in the normalization point does not affect the physical value of the electric charge needs a study of the vertex function of null-plane QED, to which the self-energy is tied by Ward-Takahashi's identities. Such a study of the vertex function, including the derivation of the electron's gyromagnetic ratio, will be done in the third part of this series.

\begin{acknowledgments}
O.A.A. thanks CAPES-Brazil for total financial support; B.M.P. thanks CNPq-Brazil for partial financial support.
\end{acknowledgments}

\appendix

\section{Normalizability of the null-plane QED$_4$}\label{sec:norm}

In this appendix we will address the problem of determining the singular order of a general transition distribution. Since it is that order which determines the number of unknown coefficients, it defines the normalization problem in CPT \footnote{The problem of renormalizability of null-plane QED in the standard approach was addressed in Refs. \cite{BrodskyRS} and \cite{Mustaki}.}. Consider, then, a causal distribution of the order $n=r+s$, which comes from the product of the transition distributions $T_r^1(x_1;\ldots;x_r)$ and $T_s^2(y_1;\ldots;y_s)$ by means of $l_p$ photon contractions and $l_f$ fermion ones; its numerical part will be:
\begin{align}\label{eq:7.1}
d&(x_1;\ldots;x_r;y_1;\ldots;y_s)\propto t_1(x_1;\ldots;x_r)\nonumber\\
&\times\prod\limits_{j=1}^{l_p}D_{ab+}(x_{r_j}-y_{s_j})\prod\limits_{m=1}^{l_f}S_+(x_{r_m}-y_{s_m})\nonumber\\
&\times t_2(y_1;\ldots;y_s).
\end{align}
In this expression, $t_1$ and $t_2$ are the numerical distributions of $T_r^1$ and $T_s^2$, respectively; $\left\{x_{r_j}\right\}$ and $\left\{y_{s_j}\right\}$ are the points at which the photon contractions take place, while $\left\{x_{r_m}\right\}$ and $\left\{y_{s_m}\right\}$ are the ones at which the fermion contractions occur. The symbol ``$\propto$'' stands instead of the equality because the ordering of the fermion's matrix commutation distributions are not taken into account. Because of translation invariance we can use relative coordinates:
\begin{align}\label{eq:7.2}
&\xi_j:=x_j-x_r\quad(j=1,\ldots, r-1),\nonumber\\
&\lambda_j:=y_j-y_s\quad(j=1,\ldots,s-1),\nonumber\\
&\lambda:=x_r-y_s,
\end{align}
so that, defining the vectors $\bm \xi=(\xi_1;\ldots;\xi_{r-1})$ and $\bm\lambda=(\lambda_1;\ldots;\lambda_{s-1})$:
\begin{align}\label{eq:7.3}
d&(\bm\xi;\bm\lambda;\lambda)\propto t_1(\bm\xi)\prod\limits_{j=1}^{l_p}D_{ab+}(\xi_{r_j}-\lambda_{s_j}+\lambda)\nonumber\\
&\times\prod\limits_{m=1}^{l_f}S_+(\xi_{r_m}-\lambda_{r_m}+\lambda)t_2(\bm\lambda).
\end{align}
As a next step we go to momentum space via Fourier's transformation:
\begin{align}\label{eq:7.4}
\hat d(\bm p;\bm q;q)\propto&\int d^{4(r-1)}\bm\xi d^{4(s-1)}\bm\lambda d^4\lambda d(\bm\xi;\bm\lambda;\lambda)\nonumber\\
&\times e^{i\bm p\bm\xi+i\bm q\bm\lambda+iq\lambda},
\end{align}
by means of which we obtain:
\begin{align}
\hat d&(\bm p;\bm q;q)\propto\int \prod_{j,m}d^4k_m d^4h_j\widehat D_{ab+}(h_j)\widehat S_+(k_m)\nonumber\\
&\quad\times\left[\int d^{4(r-1)}\bm\xi t_1(\bm\xi)e^{i\bm p\bm\xi-i\left(\sum_mk_m\xi_{r_m}+\sum_jh_j\xi_{r_j}\right)}\right]\nonumber\\
&\quad\times\left[\int d^{4(s-1)}\bm\lambda t_2(\bm\lambda)e^{i\bm q\bm\lambda-i\left(\sum_mk_m\lambda_{s_m}+\sum_jh_j\lambda_{s_j}\right)}\right]\nonumber\\
&\quad\times\left[\int d^4\lambda e^{i\left(q-\sum_mk_m-\sum_jh_j\right)}\right]\nonumber\\
&\propto \int \prod_{j,m}d^4k_m d^4h_j\widehat D_{ab+}(h_j)\widehat S_+(k_m)\hat t_1(\bm p-\bm k_r-\bm h_r)\nonumber\\
&\quad\times\hat t_2(\bm q+\bm k_s+\bm h_s)\delta\left[q-\left(\sum_jh_j+\sum_mk_m\right)\right],\label{eq:7.6}
\end{align}
with:
\begin{equation*}
(\bm p-\bm k_r-\bm h_r)_j=\left\lbrace\begin{array}{c l}
p_j-k_{r_j}-h_{r_j} & \text{if $\xi_{r_j}$ is contracted}\\
p_j & \text{if $\xi_{r_j}$ is not contracted}
\end{array}\right.
\end{equation*}
and similarly for $\bm q+\bm k_s+\bm h_s$. Now, as it was shown for Yukawa's model in Ref. \cite{APS22}, the convolution of quasiasymptotics at the $x^-$ axis generally does not exist, so that we redefine the integration variables by scaling all their four components: $\widetilde k_m=sk_m$ and $\widetilde h_j=sh_j$. Eq. \eqref{eq:7.6} then reads:
\begin{align}
\hat d&(\bm p;\bm q;q)\propto s^{-4(l_p+l_f)}\int\prod\limits_{j,m}d^4\widetilde k_md^4\widetilde h_j\widehat D_{ab+}\left(\frac{\widetilde h_j}{s}\right)\nonumber\\
&\times\widehat S_+\left(\frac{\widetilde k_m}{s}\right)\hat t_1\left(\bm p-\frac{\widetilde{\bm k_r}+\widetilde{\bm h_r}}{s}\right)\hat t_2\left(\bm q+\frac{\widetilde{\bm k_s}+\widetilde{\bm h_s}}{s}\right)\nonumber\\
&\times \delta\left[1-\frac{1}{s}\left(\sum_j\widetilde h_j+\sum_m\widetilde k_m\right)\right].\label{eq:7.7}
\end{align}
According to Sec. \ref{sec:qgi}, for all physical purposes the commutation distribution of the radiation field in Eq. \eqref{eq:7.7} can be taken as:
\begin{equation}\label{eq:7.8}
\widehat D_{ab}(p)=g_{ab}\widehat D_0(p),
\end{equation}
with $\widehat D_0(p)$ Jordan-Pauli's distribution of zero mass. Accordingly, Eq. \eqref{eq:7.7} is:
\begin{align}
\hat d_0&(\bm p;\bm q;q)\propto s^{-4(l_p+l_f)}\int\prod\limits_{j,m}d^4\widetilde k_md^4\widetilde h_j\widehat D_{0+}\left(\frac{\widetilde h_j}{s}\right)\nonumber\\
&\times \widehat S_+\left(\frac{\widetilde k_m}{s}\right)\hat t_1\left(\bm p-\frac{\widetilde{\bm k_r}+\widetilde{\bm h_r}}{s}\right)\hat t_2\left(\bm q+\frac{\widetilde{\bm k_s}+\widetilde{\bm h_s}}{s}\right)\nonumber\\
&\times\delta\left[1-\frac{1}{s}\left(\sum_j\widetilde h_j+\sum_m\widetilde k_m\right)\right].\label{eq:7.11}
\end{align}
In order to obtain the singular order of this causal distribution we need to evaluate the limit:
\begin{equation}\label{eq:7.12}
\lim_{s\to 0}s^{\omega_-}\hat d\left(\frac{\bm p}{s_*};\frac{\bm q}{s_*};\frac{q}{s_*}\right);\ \frac{p}{s_*}\equiv\left(\frac{p_+}{s};\frac{p_\perp}{s};p_-\right).
\end{equation}
Then:
\begin{align}
\hat d_0&\left(\frac{\bm p}{s_*};\frac{\bm q}{s_*};\frac{q}{s_*}\right)\propto s^{-4(l_p+l_f)}\int\prod\limits_{j,m}d^4\widetilde k_m d^4\widetilde h_j\widehat D_{0+}\left(\frac{\widetilde h_j}{s}\right)\nonumber\\
&\times\widehat S_+\left(\frac{\widetilde k_m}{s}\right)\hat t_1\left(\frac{\bm p}{s_*}-\frac{\widetilde{\bm k_r}+\widetilde{\bm h_r}}{s}\right)\hat t_2\left(\frac{\bm q}{s_*}+\frac{\widetilde{\bm k_s}+\widetilde{\bm h_s}}{s}\right)\nonumber\\
&\times\delta\left[\frac{q}{s_*}-\frac{1}{s}\left(\sum_j\widetilde h_j+\sum_m\widetilde k_m\right)\right].\label{eq:7.13}
\end{align}
In this form, by exactly the same steps given for Yukawa's model in Ref. \cite{APS22} it can be proven that:

\textbf{Lemma:} \textit{The parts of the causal distribution which do not contain the gauge terms of the commutation distribution of the radiation field have the following singular order at the $x^-$ axis:}
\begin{equation}\label{eq:7.14}
\omega_-=4-N-\frac{3}{2}M,
\end{equation}
\textit{with $N$ the number of external radiation field operators, and $M$ the number of external fermion field ones.}

We conclude that the physical $S$-operator of null-plane QED is normalizable.

Before ending this section we want to mention that, according to Eq. \eqref{eq:7.14}, the only possibilities for $\omega_-\geq 0$ are those shown in Tab. \ref{tab:nn-so-QED}.
\begin{table}[htbp]
\begin{center}
\begin{tabular}{cc|c|c}
\hline
$N$&$M$&$\omega_-$& Process\\
\hline \hline
0&0&4& Vacuum-to-vacuum distribution\\ \hline
0&2&1& Electron's self-energy \\ \hline
1&0&3& $=0$ by Furry's theorem\\ \hline
1&2&0& Vertex function\\ \hline
2&0&2& Vacuum polarization\\ \hline
3&0&1& $=0$ by Furry's theorem\\ \hline
4&0&0& Light-light scattering\\ \hline
\end{tabular}
\caption{Non-negative singular order distributions.}
\label{tab:nn-so-QED}
\end{center}
\end{table}

As we see, it is impossible to have electron selfinteracting terms, while it is, in principle, permitted to have a photon selfinteracting term of the type $\normord{A^4}$, because the singular order at the $x^-$ axis of the distributions corresponding to light-light scattering is $\omega_-=0$, so a normalization term is allowable for them. It is possible to show, however, that Ward-Takahashi's identities [Eq. \eqref{eq:8.1.4}] forbid the presence of such selfinteraction term. This makes rigorous the arguments given in Ch. 13 of Ref. \cite{JauchRohrlich}, according to which, in conventional QFT, the potential ultraviolet divergence for light-light scattering is eliminated by means of gauge invariance.

\bibliography{bibliographyAP}

\end{document}